\documentclass[preprint,review,12pt]{elsarticle}

\usepackage{amssymb}
\usepackage{amsmath}
\usepackage{amssymb}
\usepackage{sansmath}
\usepackage{newtxmath}
\usepackage[mathscr]{eucal}
\usepackage{tikz}
\usepackage{scalerel}
\usepackage{xcolor}
\usepackage{soul}

\usepackage{subfigure}
\usepackage[english]{babel}
\usepackage[T1]{fontenc}
\usepackage[utf8]{inputenc}
\usepackage{color}
\usepackage{psfrag}
\usepackage{amsmath}
\usepackage{amssymb}
\usepackage{colortbl}

\usepackage{xfrac}					
\usepackage{hyperref}
\hypersetup{colorlinks=true,allcolors=black}
\usepackage{newtxmath}
\usepackage{indentfirst}
\usepackage{float}
\usepackage[permil]{overpic}
\usepackage{bm}
\usepackage{tikz}
\usepackage{fix-cm}

\usepackage{natbib}

\journal{Journal}

\begin{document}

\begin{frontmatter}



\title{Active flow control of vertical-axis wind turbines: Insights from large-eddy simulation \\ and finite-time resolvent analysis}


\author{Lucas Feitosa de Souza} 
\author{Renato Fuzaro Miotto}
\author{William Roberto Wolf} 

\affiliation{organization={University of Campinas, Department of Energy},
            city={Campinas},
            postcode={13083-860}, 
            state={SP},
            country={Brazil}}


\begin{abstract}
Active flow control is applied to improve the aerodynamic performance of a NACA0018 airfoil operating as a single-bladed vertical axis wind turbine (VAWT). Results computed by wall-resolved large-eddy simulations (LES) highlight the detrimental effects of the dynamic stall vortex (DSV) and trailing-edge vortex (TEV) on turbine efficiency, primarily through increased drag and energy loss. The proposed flow control strategy effectively delays flow separation and suppresses large-scale vortex formation, particularly at moderate actuation frequencies. The control parameters are grounded in bi-global stability and finite-time resolvent analyses. These techniques identify the excitation of coupling modes between shear layer and wake instabilities as a mechanism for promoting flow reattachment and preventing vorticity accumulation, ultimately leading to enhanced torque production. The control strategy is energy-efficient, consuming only 1\% of the turbine’s output power while yielding substantial aerodynamic performance gains. These findings demonstrate the promise of physics-informed active flow control in mitigating dynamic stall and advancing the design of next-generation VAWTs.
\end{abstract}




\end{frontmatter}



\section{Introduction}
\label{sec1}



Vertical-axis wind turbines (VAWTs) have emerged as a promising alternative for clean energy generation, attracting increasing interest from the scientific community due to several advantages over traditional horizontal-axis wind turbines (HAWTs) \cite{hezaveh2018increasing,xie2017benefits,LEFOUEST2022505}. Unlike HAWTs, the performance of VAWTs is largely independent of wind direction, eliminating the need for yaw mechanisms and allowing a closer spacing between units in wind farms \cite{porte2013numerical}. These features make VAWTs particularly suitable for urban and complex terrain applications. In addition, they offer reduced installation and maintenance costs, due to their simpler mechanical design, fewer moving parts, and the possibility of mounting the generator at ground level. Further benefits include lower noise emissions and increased structural robustness \cite{buchner2018dynamic,rolin2018experimental,rezaeiha2018characterization,LEFOUEST2022505}.

Owing to their distinctive movement, VAWTs are susceptible to the effects of dynamic stall. This phenomenon occurs when the airfoil experiences variations in the angle of attack that surpass the static stall angle. The dynamic stall regime occurs for lifting surfaces undergoing transient motions, being characterized by significant fluctuations in aerodynamic loads, triggered by the sudden formation of a large-scale vortex at the airfoil leading edge. These loads have the potential to couple with the structural dynamics of the turbine, representing a risk of mechanical failure \cite{corke2015dynamic, BENNER2019102577}. For VAWTs, the initiation of the dynamic stall vortex (DSV) occurs in the upwind segment of the movement and is associated with blades operating at low tip-speed ratios ($\phi < 3$) \cite{10.1115/1.3268081,simao2009visualization, fujisawa2001observations,LEFOUEST2022505}. This dimensionless parameter is determined by the ratio of airfoil and freestream velocities $\phi =  \omega R  / U_\infty$, where $\omega$ is the angular velocity of the blade, $R$ is the radius of movement and $U_\infty$ is the freestream velocity. This parameter is related to the amplitude and skewness of variations in the effective inflow velocity and the blade angle of attack \cite{LEFOUEST2022505}.

The correlation between dynamic stall and aerodynamic performance of a single-bladed H-type Darrieus wind turbine was experimentally investigated by \citet{LEFOUEST2022505}. The findings of these authors indicated that a peak in the tangential force is achieved at low tip-speed ratios ($\phi < 2.5$). However, the subsequent formation of a DSV in the post-stall phase of the motion leads to a significant increase in drag and causes large load fluctuations that may compromise the structural integrity of the blade and hinder energy production. This study also emphasized that, at intermediate tip-speed ratios ($2.5 < \phi < 4$), there exists a favorable trade-off between increased tangential force and load variation, making this range optimal for the operation of such turbines.

Improving the aerodynamic performance of VAWTs still depends on mitigating the adverse effects of dynamic stall. A variety of passive and active flow control strategies have been developed to address this challenge~\cite{LEFOUEST2022505, ramos2019active, visbal_benton_control, Nathan_Webb_2018, Castaneda2022, mulleners_2023, LeFouest2024, MACPHEE2016143}. At moderate Reynolds numbers, high-frequency actuation targeting the harmonics of the laminar separation bubble near the airfoil leading edge has proven effective in suppressing the bubble bursting phenomenon responsible for DSV formation~\cite{visbal_benton_control, Visbal2023_cavity}. In transitional Reynolds numbers, on the other hand, numerical and experimental studies have shown that moderate-frequency suction and blowing can effectively suppress flow separation and sustain suction near the leading edge~\cite{ramos2019active, Castaneda2022, Nathan_Webb_2018, desouza2025dutycycleactuationdragreduction}. These findings are particularly relevant in the context of VAWTs, as most configurations operate in the transitional flow regime~\cite{simao2009visualization, LEFOUEST2022505}.

When considering flow control strategies at transitional Reynolds numbers, the key mechanism behind the suppression of flow separation lies in the manipulation of Kelvin--Helmholtz instabilities within the leading-edge shear layer. However, there is still limited understanding of why certain frequency ranges of actuation yield better results. In this context, local stability analysis conducted by \citet{gupta_ansell_orr_sommerfeld_2020} demonstrated that, for small variations in angle of attack, the dispersion of the eigenvalue spectrum for shear layer profiles remains small, even when accounting for the time-dependent base flow of an airfoil undergoing dynamic stall. Building on this, \citet{desouza2025dutycycleactuationdragreduction} showed that, for a periodically plunging airfoil, information about the most unstable eigenvalue near the instant of minimal angle-of-attack variation, obtained by a local linear stability analysis of the shear layer, can be used to guide both the frequency and duty cycle of a suction and blowing jet actuator. This approach led to a significant reduction in both drag and actuation effort.

Thick airfoil profiles are typically employed in VAWTs and trailing-edge stall is more likely to occur. In such cases, dynamic stall is initiated by a large region of flow reversal that develops from the trailing edge toward the leading edge~\cite{LEFOUEST2022505, simao2009visualization}. This flow behavior, combined with the complex motion kinematics of VAWTs, makes it challenging to define a representative position for conducting a local stability analysis. As a result, bi-global stability and resolvent analyses are more appropriate methods for the study of VAWTs. These techniques should also provide insights in terms of flow control strategies for dynamic stall mitigation, including actuator placement and frequency.

For time-invariant base flows, \citet{yeh_taira_2019} demonstrated the utility of bi-global resolvent analysis, and its discounted variant, for gaining physical information of the forced response of an airfoil undergoing static stall. Their evaluation of the forcing and response resolvent modes, as well as the associated gains, motivated the development of effective control strategies to mitigate flow separation and improve aerodynamic performance in terms of lift-to-drag ratio. Furthermore, \citet{Ribeiro_Taira_2024} showed that tri-global resolvent analysis can be applied to infer control strategies aimed at enhancing the aerodynamic performance of low-aspect-ratio wings.

Motivated by the demonstrated efficacy of stability-based approaches for flow control, we employ bi-global resolvent analysis informed by wall-resolved large-eddy simulations (LES) to investigate and mitigate dynamic stall in VAWTs. Unlike the majority of recent numerical studies, which predominantly adopt RANS methodologies~\cite{firdaus2015numerical,liu2022aerodynamic,aboelezz2022novel,hao2022study,wang_2024,MACPHEE2016143,MCNAUGHTON2014124, ALMOHAMMADI2015144}, our high-fidelity LES framework resolves the unsteady boundary layer dynamics critical to the inception and evolution of the DSV. The resolvent analysis, performed about a phase-averaged base flow extracted from LES, reveals the spatial structure and frequency content of the most amplified flow disturbances, enabling a physics-based strategy for actuator placement and control input design. This integrated approach facilitates targeted suppression of flow separation and results in improved aerodynamic performance for a rotating airfoil subjected to VAWT-relevant kinematics.


\section{Theoretical and numerical methodology}
\label{sec2}

\subsection{Large-eddy simulations}

Wall-resolved LES are performed on a single NACA0018 airfoil under the characteristic motion of a VAWT by solving the compressible form of the Navier-Stokes equations. The equations are solved in non-dimensional form, in a non-inertial frame of reference, and further details are presented in \ref{appendix}. The motion effects are accounted for by reevaluating the right-hand side of the momentum and total energy equations at each timestep. This approach does not involve overhead as the computational grid can be kept fixed for any displacements of the airfoil.


An O-type grid is employed and the governing equations are solved in a general curvilinear coordinate system. The spatial discretization of the flow equations is performed using the sixth-order accurate compact finite-difference scheme from \citet{Nagarajan2003}. This methodology solves the advective and viscous fluxes using a staggered grid setup. The sixth-order compact interpolation scheme presented in the same reference is used to interpolate flow quantities in the staggered grid approach. No explicit subgrid scale (SGS) model is applied. However, to preserve stability of the numerical simulation, a sixth-order compact filter \cite{lele1992compact} is applied to control high-wavenumber numerical instabilities that may arise from mesh curvature, numerical interpolation and boundary conditions. The present high-order compact filter serves as an approximation to an SGS model as discussed in Ref. \cite{mathew2003explicit}.

The time integration of the flow equations is performed using an explicit third-order compact-storage Runge-Kutta scheme in regions away from solid boundaries. Near the airfoil surface, an implicit second-order scheme is applied to overcome the stiffness problem typical of boundary layer grids. Sponge layers and characteristic boundary conditions based on Riemann invariants are applied in the farfield, and adiabatic no-slip boundary conditions are used at the airfoil surfaces. Periodic boundary conditions are enforced along the span. The present numerical tool has been validated against experimental results and high-fidelity numerical simulations of turbulent flows \cite{Nagarajan2003, Bhaskaran, wolf2012}, besides the study of dynamic stall \cite{ramos2019active, miotto2022analysis, miotto2021pitch}. A study regarding the grid convergence and the effects of spanwise domain extent for the present simulation can be found in Ref. \cite{VAWT_validation_AIAA}.

\subsection{Flow setup}

\begin{figure}[H]
\centering
\includegraphics[trim = {0cm 0 0cm 0cm}, clip, width= 0.99\textwidth]{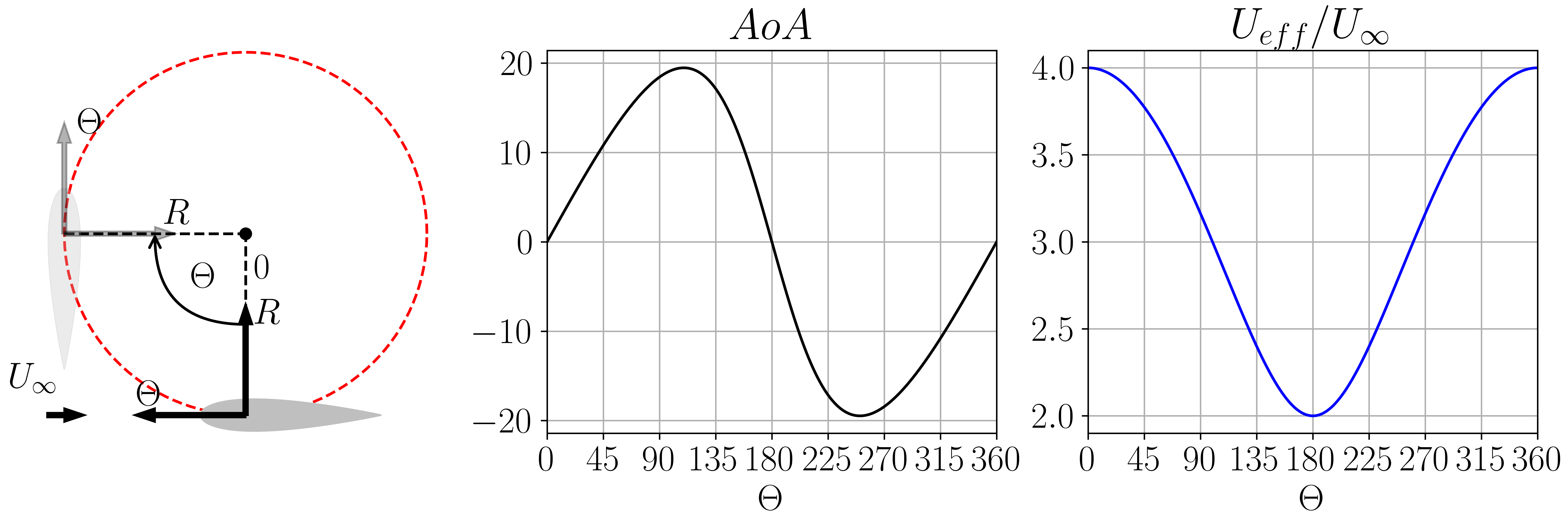}\hfill
\caption{Sketch describing the rotational motion of the single bladed VAWT highlighting the reference system used to calculate the aerodynamic coefficients (left), variation of the effective angle of attack during one revolution of the airfoil for $\phi = 3$ (center), variation of the effective speed experienced by the airfoil during the cycle (right).}
\label{fig:motion}
\end{figure}

We consider a single NACA0018 airfoil undergoing the typical circular motion of VAWTs, as shown in Fig. \ref{fig:motion}. The simulations are performed with the airfoil motion beginning at the topmost position, but the first half cycle is discarded due to initial transients. Results are reported with the starting point being the bottommost position ($\Theta = 0$). 

In a vertical-axis wind turbine, the rotation  of the blades about an axis perpendicular to the freestream causes periodic changes in the effective angle of attack and inflow velocity during each cycle. In this context the blade kinematics is governed by the tip-speed ratio $\phi = \omega R / U_{\infty}$, which is the ratio between the blade rotational speed to the freestream velocity \citep{fujisawa2001observations, LEFOUEST2022505, 10.1115/1.3268081, simao2009visualization}. Starting from an initial position facing the wind, the variation in the (geometric) effective angle of attack ($AoA_{\text{eff}}$) and the effective velocity ($U_{\text{eff}}$) can be described as a function of the blade azimuthal position $\Theta$ and tip speed ratio as:

\begin{equation}
    AoA_{\text{eff}} = \arctan \left (\frac{\sin \Theta}{\phi + \cos{\Theta}} \right ) \mbox{ , and}
\end{equation}
\begin{equation}
    \frac{U_{\text{eff}}}{U_{\infty}} = \sqrt{1 + 2 \phi \cos (\Theta) + \phi^2} \mbox{ .}
\end{equation}

In our simulations, we maintain a constant Reynolds number of $\text{Re}_{\omega}= \omega R c/\nu= 50,000$, based on the tip speed, and the tip speed ratio and rotation radius are $\phi = 3$ and $R/c = 2.5$, respectively. The present investigation will focus on a blade operating at a tip-speed ratio of \(\phi = 3\), as this regime has been identified as favorable for the application of small-scale turbines, according to \citet{LEFOUEST2022505}. In this configuration, variations in the effective angle of attack and velocity are illustrated in Fig. \ref{fig:motion}. At the starting position, \(\Theta = 0\), the turbine faces the freestream with an effective angle of attack of 0 degrees. The angle of attack then gradually increases, eventually surpassing the static stall angle. This occurs during the upwind portion of the rotation, between \(0 \leq \Theta \leq 90\). As a result, vortices form on the suction side of the blade, which negatively impacts the aerodynamic performance of the turbine in the critical energy-production segment \((45 < \Theta < 180)\).

\subsection{Linear stability and resolvent analyses}

To perform a biglobal resolvent analysis, we employ the MATLAB package \textit{LinStab2D} developed by \citet{Martini2024}. This computational tool facilitates the implementation of temporal stability and resolvent analyses for compressible viscous flows in curvilinear grids. The mechanisms of generation and amplification of flow instabilities in a finite time window can be explained by linear stability theory. Hence, a bi-global  discounted resolvent analysis \cite{yeh_taira_2019, Yeh_taira_2020, Rolandi2024} will be employed in this work to investigate the most unstable frequencies and their receptivity to disturbances in the onset process of dynamic stall. Through a Reynolds decomposition, it is possible to split the unsteady flow $\boldsymbol{q}(\boldsymbol{x},t)$ into a base (mean) flow $\bar{\boldsymbol{q}}(\boldsymbol{x})$ plus a time-dependent fluctuation component $\boldsymbol{q}'(\boldsymbol{x},t)$. Here, $\boldsymbol{q} = [\rho, u, v, w, T]$. If the fluctuations are sufficiently small, the Navier-Stokes equations can be linearized about the base flow.

Although it is possible to consider the turbulent mean flow $\bar{\boldsymbol{q}}$ as the base flow, the linear  analysis would not hold because such state is not an equilibrium point of the governing equations. Nonetheless, the use of a time-averaged base flow may provide some insights as a model \cite{Taira_AIAAJ2017,Taira_AIAAJ2020, Rolandi2024}. In this context, for statistically stationary flows, the base flows are usually computed by time averaging the instantaneous solutions. However, for an airfoil undergoing dynamic stall, this averaging process does not provide an approximation to a suitable base flow. Hence, we employ a 6-cycle phase-averaged solution $\bar{q}_{\Theta}$ immediately prior to the onset of dynamic stall. By doing so, we are neglecting the temporal evolution of the baseflow with respect to the timescales of the disturbances. It is important to mention that the timescales of the turbine motion are slower than those of instabilities.
In this case, with the addition of a time-dependent external forcing $\boldsymbol{f}'(\boldsymbol{x},t)$, the linearized Navier-Stokes equations (LNS) can be written as
	\begin{equation}
	\frac{\partial \boldsymbol{q}'}{\partial t} = \boldsymbol{\mathcal{L}} \boldsymbol{q}' + \boldsymbol{\mathcal{B}} \boldsymbol{f}' \mbox{ ,}
	\label{eq:LNS_1}
	\end{equation}
where $\boldsymbol{\mathcal{B}} = \boldsymbol{\mathcal{B}}(\boldsymbol{x})$ is an operator which serves as a spatial window where forcing is applied. The matrix $\boldsymbol{\mathcal{L}} = \boldsymbol{\mathcal{L}}(\bar{\boldsymbol{q}}_\Theta)$ is the bi-global LNS operator with the spanwise phase-averaged flow properties $\overline{\boldsymbol{q}}_{\Theta}(\boldsymbol{x})$ as base flow.
	
The evolution of linear disturbances by equation \ref{eq:LNS_1} can be performed by a direct analysis of the operator $\boldsymbol{\mathcal{L}}$ or time-integrating the disturbances in a linearized version of the CFD code. In the former case, with a transformation
	\begin{equation}
	\bullet' (x,y,z,t) = \frac{1}{2 \pi \mathrm{i}}\int_{\psi -\mathrm{i}\infty}^{\psi + \mathrm{i}\infty} \int_{-\infty}^{\infty} \hat{\bullet}(x,y,\beta,\text{St}) e^{\mathrm{i} \beta z +s t} \, d \beta \, d s \mbox{ ,}
	\end{equation}
where the wavenumber $\beta \in \mathbb{R}$ and the Laplace variable $s \in \mathbb{C}$, it is possible to write the forced LNS equations in discrete form as 
	\begin{equation}
	s \hat{\boldsymbol{q}} = \mathsf{L} \hat{\boldsymbol{q}} + \mathsf{B} \hat{\boldsymbol{f}} \mbox{ .}
	\label{eq:discrete_FNS}
    \end{equation}
In this case, the linear operator becomes a function of the spanwise wavenumber $\mathsf{L} = \mathsf{L}(\bar{\boldsymbol{q}_{\Theta}},\beta)$ and $\mathsf{B}$ is the discrete version of $\boldsymbol{\mathcal{B}}$. If forcing $\hat{\boldsymbol{f}}$ is absent, the LNS equations can be analyzed separately for each wavenumber $\beta$ as an eigenvalue problem (linear stability analysis) as
	\begin{equation}
	\mathsf{Q} \bm{\Lambda} = \mathsf{L} \mathsf{Q} \mbox{ .}
	\label{eq:LSA}
	\end{equation}
    
%
In case forcing is applied at a angular frequency $2\pi \text{St}$, and for a finite-time horizon of analysis $1/\psi$ equation \ref{eq:discrete_FNS} yields
	\begin{equation}
	\hat{\boldsymbol{q}} = ((\mathrm{i} 2\pi \text{St} + \psi) \mathsf{I} - \mathsf{L} )^{-1} \mathsf{B} \hat{\boldsymbol{f}} = \mathsf{H} \, \mathsf{B} \hat{\boldsymbol{f}} \mbox{ ,}
	\label{eq:resolvent_1}
	\end{equation}
where the matrix $\mathsf{H} = \mathsf{H}(\bar{\boldsymbol{q}_\Theta},\beta, \text{St}, \psi)$ is the discrete discounted resolvent operator and can be expressed as \cite{yeh_taira_2019, Yeh_taira_2020}:

\begin{equation}
    \mathsf{H} = \left[ \mathrm{i}  2\pi \text{St} \mathsf{I} - (\mathsf{L} -  \psi\mathsf{I}) \right]^{-1} \mbox{ ,}
\end{equation}
Here, setting $\psi = 0$ retrieves the standard resolvent operator \cite{reddy_henningson_1993,Schmid2007annrev,mckeon_sharma_2010,Taira_AIAAJ2017,Taira_AIAAJ2020}.	

For non-normal operators, eigenvalue sensitivity and energy amplification are related to the induced $L_2$ norm of operator $\mathsf{H}$, \textit{i.e.}, the leading singular value $\sigma$ obtained from singular value decomposition
	\begin{equation}
	\mathsf{H} = \mathsf{U} \bm{\Sigma} \mathsf{V}^{*} \mbox{ .}
	\label{eq:svd_resolvent}
	\end{equation} 
In this equation, $\mathsf{V}$ and $\mathsf{U}$ are unitary matrices holding right and left singular vectors and $\bm{\Sigma}$ is a diagonal matrix containing the singular values $\sigma$, such that
	\begin{equation}
	\left\| \mathsf{H} \right\| = \max( \sigma ) \; = \sigma_1 \mbox{ .}
	\end{equation}
The first column of $\mathsf{V}$ contains the forcing term that produces the largest response in the flow (first column in matrix $\mathsf{U}$) with the amplification ratio given by $\bm{\Sigma}$.
	
A transformation using matrix $\mathsf{W}$ is applied to convert variables $\hat{\boldsymbol{q}}$ and $\hat{\boldsymbol{f}}$ with an appropriate energy norm prior to performing the SVD. In the case of compressible flows, the Chu norm \cite{ChuNorm}, which relates density, velocity, and temperature, is used. A spatial window $\mathsf{C} = \mathsf{C}(\boldsymbol{x})$ is applied to limit the domain of analysis and the system response as
	\begin{equation}
	\hat{\boldsymbol{y}} = \mathsf{W} \mathsf{C} \hat{\boldsymbol{q}} \mbox{ .}
	\label{eq:resolvent_2}
	\end{equation}
Combining equations \ref{eq:resolvent_1} and \ref{eq:resolvent_2} leads to the modified resolvent operator
	\begin{equation}
	\mathsf{H}_w = \mathsf{W} \mathsf{C} \left[ \mathrm{i}  2\pi \text{St} \mathsf{I} - (\mathsf{L} -  \psi\mathsf{I}) \right]^{-1} \mathsf{B} \mathsf{W}^{-1} \mbox{ .}
	\label{eq:final_resolvent}
	\end{equation}
	
The amplification mechanisms of flow disturbances can be identified by using the eigenvalue decomposition of the discounted operator $\mathsf{L}_\psi =  (\mathsf{L} -  \psi\mathsf{I})$ in the resolvent, yielding
	\begin{equation}
	\mathsf{H}_w = \mathsf{W} \mathsf{C} (\mathrm{i} 2\pi \text{St} \mathsf{I} - \mathsf{Q}_\psi \bm{\Lambda}_{\psi} \mathsf{Q}_\psi^{-1} )^{-1} \mathsf{B} \mathsf{W}^{-1} = \mathsf{W} \mathsf{C} \mathsf{Q}_\psi \left(\mathrm{i} 2\pi \text{St} \mathsf{I} - \bm{\Lambda}_\psi \right)^{-1} \mathsf{Q}_\psi^{-1} \mathsf{B} \mathsf{W}^{-1} \mbox{ ,}
	\label{eq:resolvent_eigen}
	\end{equation}
where $\bm{\Lambda}_\psi$ and $\mathsf{Q}_\psi$ are the eigenvalues and eigenvectors of $\mathsf{L}_\psi$, respectively. The bounds of the induced $L_2$ norm are

\begin{equation}
	\left\| \left( \mathrm{i} 2\pi \text{St} \mathsf{I} - \bm{\Lambda}_\psi \right)^{-1} \right\|
	\le \left\| \mathsf{H}_w \right\|
	\le \left\| \left( \mathrm{i} 2\pi \text{St} \mathsf{I} - \bm{\Lambda}_\psi \right)^{-1} \right\| \; || \mathsf{Q}_\psi || \; || \mathsf{Q}_\psi^{-1} || \mbox{ ,}
\label{eq:ressonance}
\end{equation}
where the lower bound is the case of an operator with orthonormal eigenvectors. In this case, the norm depends only of the resonances, following a $\sfrac{1}{R}$ decay based on the distance $R$ in the complex plane with respect to the eigenvalues. For non-normal systems, the norm also depends on the pseudoresonances, measured by the product of eigenvectors $\mathsf{Q_\psi}$ and their inverse $\mathsf{Q_\psi}^{-1}$. Note that the spatial support of the eigenvectors is also a function of the finite-time horizon considered, representing the spatial dynamics until a finite time $t =1/\psi$. The weighting $\mathsf{W}$ and the two windowing matrices $\mathsf{B}$ and $\mathsf{C}$ are included in the norm calculation (see \cite{schmidYellowBook,McKeon2018} for more details).

From equation~\ref{eq:ressonance}, it can be observed that for a marginally stable system—i.e., a system whose eigenvalue is given by $\Lambda(\mathsf{L}) = \mathrm{i} \lambda_i + \lambda_r$, with $\lambda_r \to 0$ and $\lambda_r \leq 0$, the maximum resonance occurs in the limit $\psi \to 0$ and $2\pi \text{St} = \lambda_i$, consistent with the assumptions made in a standard resolvent analysis, where the growth of instabilities in an asymptotic time is considered. However, in the case of an unstable system, the resonances in the resolvent response must be identified in the limit $\psi \to \max(\lambda_r)$ and $2\pi \text{St} \to \lambda_i$. This is seen in the analogous transfer function $\text{H}(\text{St},\psi)$ for a single eigenvalue $\lambda$.
\begin{equation}
\left\lVert \text{H}(\text{St},\psi ) \right\rVert = \left\lVert \frac{1}{\mathrm{i}( 2\pi \text{St} -  \lambda_i) + (\psi - \lambda_r)} \right\rVert \mbox{ .}
\label{eq:analogoustf}
\end{equation}
Therefore, to conduct a discounted resolvent analysis, it is essential to first identify the most unstable eigenvalue of the discrete Navier–Stokes operator. This is achieved by performing a linear stability analysis, which involves solving the eigenvalue problem associated with the equation~\ref{eq:LSA}. 

Although stability analysis already provides valuable information on the unsteady frequencies of a system, the application of the discounted resolvent analysis offers significant advantages, particularly in the context of flow control for complex systems where finite-time dynamics are of interest. Physically, the discounting approach can be interpreted as an input–output analysis over a finite-time horizon, where there exists a time \( t \) at which the transient energy amplification of an unstable system coincides with the long-term amplification of the corresponding discounted system~\cite{Yeh_taira_2020, jovanovic_thesis}. The associated forcing and response modes derived from this framework yield critical insights for the development of effective control strategies, including optimal actuator placement, and facilitate a deeper understanding of the physical mechanisms responsible for control efficacy.

	
To perform the resolvent (pseudospectral) analysis, the discrete linear operator is computed using the second-order accurate methodology from Ref. \cite{Martini2024}. As discussed above, the base flow is computed by the spanwise and phase-averaged solution from the LES, $\bar{\boldsymbol{q}_{\Theta}} = [\bar{\rho}, \bar{u}, \bar{v}, \bar{w}, \bar{T}]$. For the far-field and airfoil surface, Dirichlet boundary conditions are set for all variables $[\rho' , u' , v' , w' , T'] = [0, 0, 0, 0, 0]$. With these boundary conditions and the base flow $\bar{\boldsymbol{q}_\Theta}$, the linear operator can be computed in its discrete form $L(\bar{\boldsymbol{q}}_{\Theta},\beta)$ for a prescribed spanwise wavenumber $\beta$. Here, we use $\beta = 0$ since 2D actuation was shown to be effective for dynamic stall at transitional Reynolds numbers \cite{ramos2019active, desouza2025dutycycleactuationdragreduction, Visbal2023_cavity}. Finally, the SVD in equation \ref{eq:final_resolvent} is performed to uncover the optimal forcing frequencies and the finite-time response of the flow system. 

A grid convergence study was conducted to ensure accuracy of the linear stability analysis results. The mesh resolution was determined based on the unstable mode with the highest frequency identified in the spectrum, such that the spatial discretization error remained below 1\% within the regions of significant spatial support of the corresponding unstable eigenvectors. This criterion was established by evaluating the modified wavenumber associated with the finite-difference scheme employed and also by verifying the convergence of the eigenvalue spectrum.

\section{Results}
\label{sec3}

This section presents results of a single-bladed VAWT for uncontrolled (baseline) and controlled flows at different frequencies obtained by wall-resolved LES. In order to gain a deeper understanding of the physical processes related to flow actuation, bi-global linear stability and resolvent analyses are also performed. The latter approach is applied to analyze the evolution of flow disturbances in a finite-time sense, and it helps elucidate the effectiveness of flow actuation at different frequencies. 

To evaluate the aerodynamic performance of a VAWT, the primary metric is the torque generated by the blade relative to the axis of rotation. When defining clockwise rotation (in the direction of motion) as positive, a positive torque contributes to energy production. The aerodynamic torque is calculated in non-dimensional form as follows
\begin{equation}
    \vec{\tau} = \frac{1}{0.5  \rho_{\infty} (\omega R)^2 c^2}  \oint \Vec{r}(s) \times \Vec{f}(s) ds \mbox{ ,}
\end{equation}\label{eq : tau}
where $s$ is the position along the airfoil surface, $\Vec{r}(s) = (r_x(s) - r_{x_o}, r_y(s) - r_{y_o}, 0)$ is the lever from the airfoil surface about the tubine axis of rotation and $\vec{f}(s) = (f_x(s), f_y(s), 0)$ represents the aerodynamic load at a position $s$. Using the data in the non-inertial reference frame, as shown in Fig. \ref{fig:motion}, we can also calculate the non-dimensional aerodynamic loads in the $\Theta$ and $R$ directions as 
\begin{equation}
    F_{\Theta} = \frac{1}{0.5 \rho_{\infty} (\omega R)^2 c} \oint f_{\Theta}(s) ds  \quad\text{and}\quad   F_{R} = \frac{1}{0.5 \rho_{\infty} (\omega R)^2 c} \oint f_{R} (s) ds \mbox{ .}
\end{equation}

In this frame of reference, a negative $F_{\Theta}$ corresponds to aerodynamic drag that opposes the blade motion responsible for energy production, making it a critical quantity for performance evaluation. Although forces in the radial direction also contribute to the aerodynamic torque, their influence is secondary compared to the dominant effect of forces in the $\Theta$-direction.

\subsection{Overall features of the baseline and actuated turbines}
\label{sec:flow_features}
%

\begin{figure}[H]
    \centering
    \begin{overpic}[trim = 2cm 0cm 2cm 0cm,clip,width=0.51\textwidth]{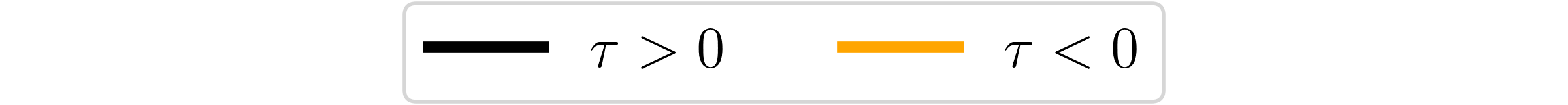}
    \end{overpic}
    \begin{overpic}[trim = 1.2cm 0cm 1.8cm 0cm,clip,width=0.49\textwidth]{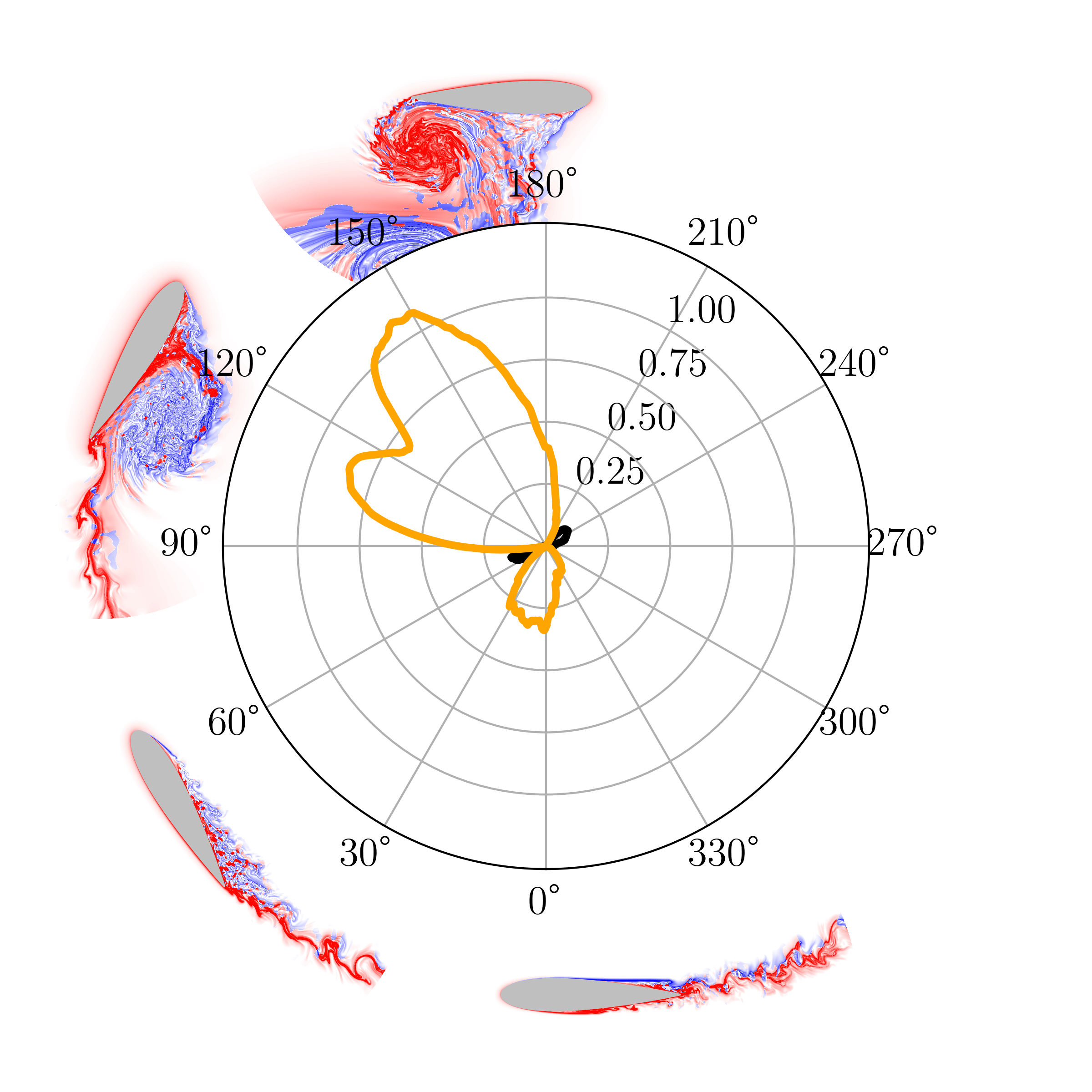}
    \put(60,900){a)}
    \end{overpic}
    \begin{overpic}[trim = 1.2cm 0cm 1.8cm 0cm,clip,width=0.49\textwidth]{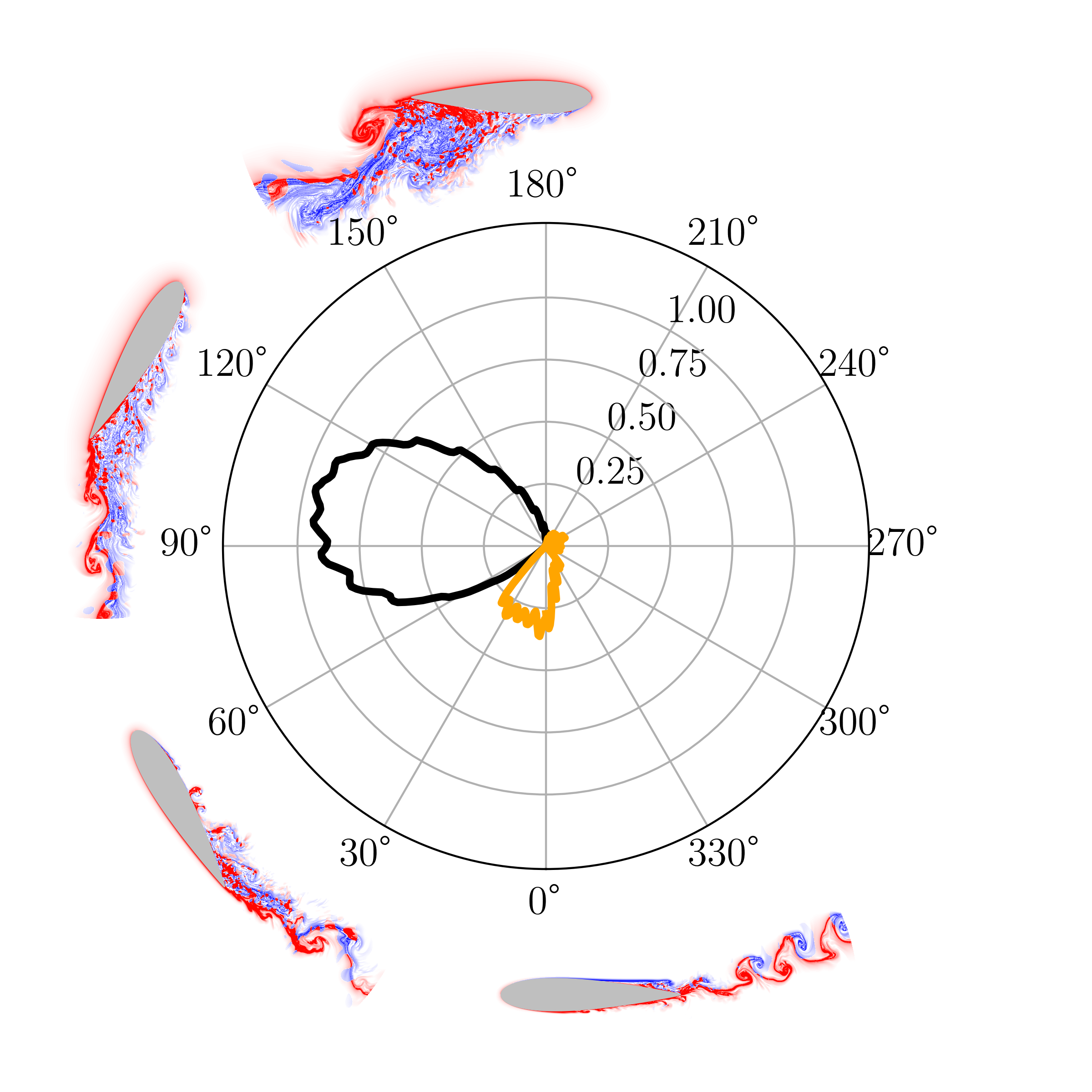}
    \put(60,900){b)}
    \end{overpic}
    \begin{overpic}[trim = 0cm 0cm 0cm 0cm,clip,width=0.35\textwidth]{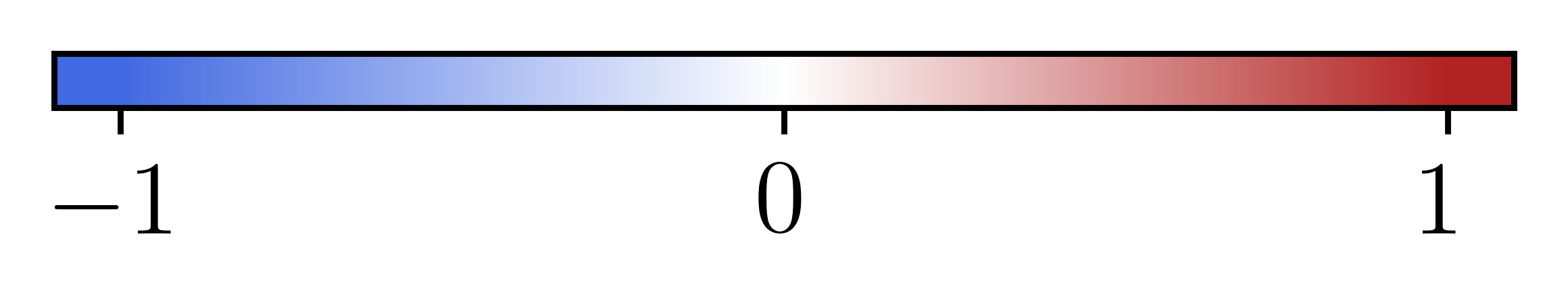}
    \put(375,200){$\text{FTLE}_{\Omega_z}$}
    \end{overpic} 
\caption{Contours of $\text{FTLE}_{\Omega_{z}}$ illustrating the evolution of the flow field at various positions during the upwind motion in the turbine cycle. At the center, a polar plot depicts the magnitude of the torque coefficient. The color scheme is designed to highlight regions contributing positively to energy production (black) and those that impede energy production (orange). Figures (a) and (b) present the results for the baseline (uncontrolled) and controlled cases with actuation frequency $\text{St} = 8$, respectively. A movie provided as supplementary material shows the flow field evolution in greater detail.}
\label{fig:cycles}
\end{figure}

This section describes the main flow characteristics of a single-bladed vertical-axis wind turbine operating at a Reynolds number of $ Re_{\omega} = 50,000$ and a tip-speed ratio of $ \phi = 3 $. To simulate a realistic, quasi-incompressible flow environment, the freestream Mach number is set to $ M = 0.025 $. Figure~\ref{fig:cycles} shows the flow field around the airfoil at four different azimuthal positions during the upwind portion of the turbine cycle, comparing the baseline (Fig.\ref{fig:cycles}(a)) and the controlled case (Fig.\ref{fig:cycles}(b)). 
The control parameters used in the previous figure correspond to the most effective configuration among all tested cases in terms of suppressing flow separation and enhancing aerodynamic performance. Importantly, the selection of these parameters is not arbitrary; it is grounded in insights gained from bi-global stability and resolvent analyses, which identify favorable frequency ranges for effective flow control. These analyses and their role in guiding the control strategy are discussed in detail in a later section of the manuscript. 

The central polar plot in each frame illustrates the aerodynamic torque generated over the cycle, with color indicating its sign. In these plots, a positive (clockwise) torque, shown in black, contributes to energy production. The analysis of the torque plot reveals that the main events affecting the aerodynamic performance of the VAWT occur during the upwind portion of the cycle ($ 0^\circ \leq \Theta \leq 180^\circ$). Hence, our analysis will focus exclusively on this phase of motion. In the same figure, contour plots of the finite-time Lyapunov exponent colored by the sign of vorticity $\text{FTLE}_{\Omega_z}$ allow for a visualization of the development of the dynamic stall, highlighting how variations in the effective angle of attack lead to the formation of coherent flow structures. In these contours, the red (blue) color indicates a positive (negative) vorticity. The present Lagrangian method is chosen because it provides a direct measure of material transport in forward time, closely mimicking experimental flow visualizations using tracers. This approach is particularly suitable for our purposes, as it allows us to observe the interactions between vortices, shear layers, and wake patterns associated with the onset and evolution of dynamic stall \cite{HALLER2000352}. A movie is also provided as supplementary material to show the flow field evolution in greater detail.


Initially, the turbine is positioned at \(\Theta = 0^\circ\), directly facing the freestream. For both the uncontrolled and controlled cases, a laminar boundary layer develops along the airfoil, with a small section of separated flow downstream due to a shear layer developing from the mid-chord of the airfoil. In the portion of the cycle for $0^\circ<\Theta<30^\circ$, we can see that the aerodynamic torque is negative in both cases, due to the drag forces the blade is subjected to, which can be attributed to the higher effective speed. As the uncontrolled turbine moves in the cycle between \(30^\circ \leq \Theta \leq 60^\circ\), Kelvin-Helmholtz instabilities develop in the shear layer, leading to the accumulation of negative vorticity (blue colors in the contour) in the aft portion of the airfoil. The increase in the effective angle of attack improves suction on the leading edge, briefly reversing the sign of the torque. During the short interval of \(60^\circ < \Theta < 90^\circ\), the aerodynamic forces act favorably, contributing to a small energy production shown by the black line in Fig. \ref{fig:cycles}(a).

For $\Theta > 90^\circ$ the process of coalescence of Kelvin-Helmhotz instabilities leads to the formation of a DSV in the uncontrolled case. This vortex grows fed by the shear layer emanating from the leading edge, giving rise to the large coherent structure seen at $\Theta = 120^\circ$. The DSV develops near the trailing edge and induces a large aerodynamic drag, drastically reducing energy production, as can be seen by the first negative torque peak in Fig. \ref{fig:cycles}(a). A second peak of negative torque can also be seen at $\Theta \approx 150^\circ$. This peak is related to the formation of a trailing-edge vortex (TEV) that interacts with the blade reducing the surface pressure at the trailing edge. 

A comparison between Figs. \ref{fig:cycles}(a) and (b) highlights the lost potential for energy extraction caused by dynamic stall. In the baseline case, the presence of large-scale coherent structures leads to a higher aerodynamic drag, which significantly reduces energy production. In contrast, the controlled case demonstrates that actuation at a non-dimensional frequency of $\text{St} =fc/U_\infty= 8$ effectively suppresses these coherent vortices, resulting in an extended period of useful torque generation. At the beginning of the cycle, near \( \Theta = 0^\circ \), the boundary layer behavior is similar in both cases. However, a notable difference appears in the level of organization of the Von Kármán wake pattern, which is more structured in the controlled case, indicating a potential coupling mechanism between the actuation frequency and the wake dynamics. This effect throughout the cycle will be further explored in the following sections to discuss the effectiveness of this particular actuation frequency.

As the blade advances through the upwind portion of the cycle, the controlled case exhibits the ejection of discrete vortices that locally promote flow reattachment and reduce the amount of negative vorticity near the surface. However, at this stage, there is still no significant improvement in torque production. Aerodynamic enhancement becomes evident in the interval \( 60^\circ \leq \Theta \leq 150^\circ \). In particular, the absence of large-scale coherent structures such as the DSV and TEV stands out. Without the additional drag caused by the blade interaction with these structures, the turbine generates a net positive torque.

\subsection{Impact of actuation on aerodynamics}

As demonstrated in the previous section, dynamic stall has a detrimental impact on the aerodynamic performance of VAWTs. The aerodynamic drag resulting from the interaction of the blades with the DSV and the TEV hinders the production of useful energy. Consequently, advances in the development of efficient turbines depend on the effective reduction of drag induced by dynamic stall. This section focuses on the application of active flow control of the present baseline case, which characterizes the typical operating regime for VAWTs.

Suction and blowing actuation represents a promising alternative for mitigating the undesirable effects of dynamic stall. By introducing disturbances into the boundary layer at an appropriate frequency, this technique can effectively reduce flow separation on the suction side of the blade and promote sustained suction near the leading edge \cite{Castaneda2022, ramos2019active, Nathan_Webb_2018}. Furthermore, flow control influences the accumulation of vorticity generated by the coalescence of Kelvin-Helmholtz instabilities that develop within the shear layer originating from the separation region. In this work, we employ a similar leading-edge jet actuation as that described by \citet{ramos2019active}. The control function used is described by the equation \ref{eq:control}. 

\begin{equation}
    U_{\text{jet}} (s,t) = {U_{\infty}}  \exp\left({\frac{-(s^* - 0.01)^2}{4.5}}\right) \sin{(\text{St} 2 \pi t)} \mbox{ .} 
\end{equation}
\label{eq:control}
Here, the jet actuation is applied as a spatial Gaussian distribution at the leading edge and it has a maximum speed of $U_{\text{jet-max}}$. The frequency of actuation is given in terms of the Strouhal number $\text{St}$. The spatial support of the actuation is given by $s^*$ and comprises 2\% of the airfoil chord.

To quantify and compare the work generated by the turbine during a complete rotation cycle ($W_{\tau}$) with the work expended by the actuation jet ($W_{\text{jet}}$), the ratio of work coefficients is evaluated. The definition of $W_\text{jet}$ is given by:
\begin{equation}
W_{\text{jet}} = \frac{1}{0.5 \rho_{\infty} \phi^3 U_{\infty}^3 L_z \Delta s T_c} \int_{0}^{T_c} \int_{S_0}^{S_n} \int_{0}^{z} \rho_{\infty} \lvert U_{\text{jet}}^3 (s,z,t) \rvert \, ds\, dz\, dt \mbox{ ,}
\end{equation}
where \( \rho_\infty \) is the freestream density, \( U_\infty \) is the freestream velocity, \( L_z \) is the spanwise length, $\Delta s$ is the streamwise support of the actuation, \( T_c \) is the period of a rotation cycle, and \( U_{\text{jet}}(s,z,t) \) represents the jet velocity.
For the turbine work coefficient, the resulting expression is:
\[
W_{\tau} =  \frac{1}{\pi \phi^2 \rho_{\infty} U_{\infty}^2 c^3} \int_{0}^{\Theta} \tau(\Theta) \, d\Theta \mbox{ ,}
\]
where $\Theta$ is given in radians. 
%

\begin{figure}[H]
    \centering
    \begin{overpic}[trim = 0cm 0.0cm 0cm 0cm,clip,width=\textwidth]{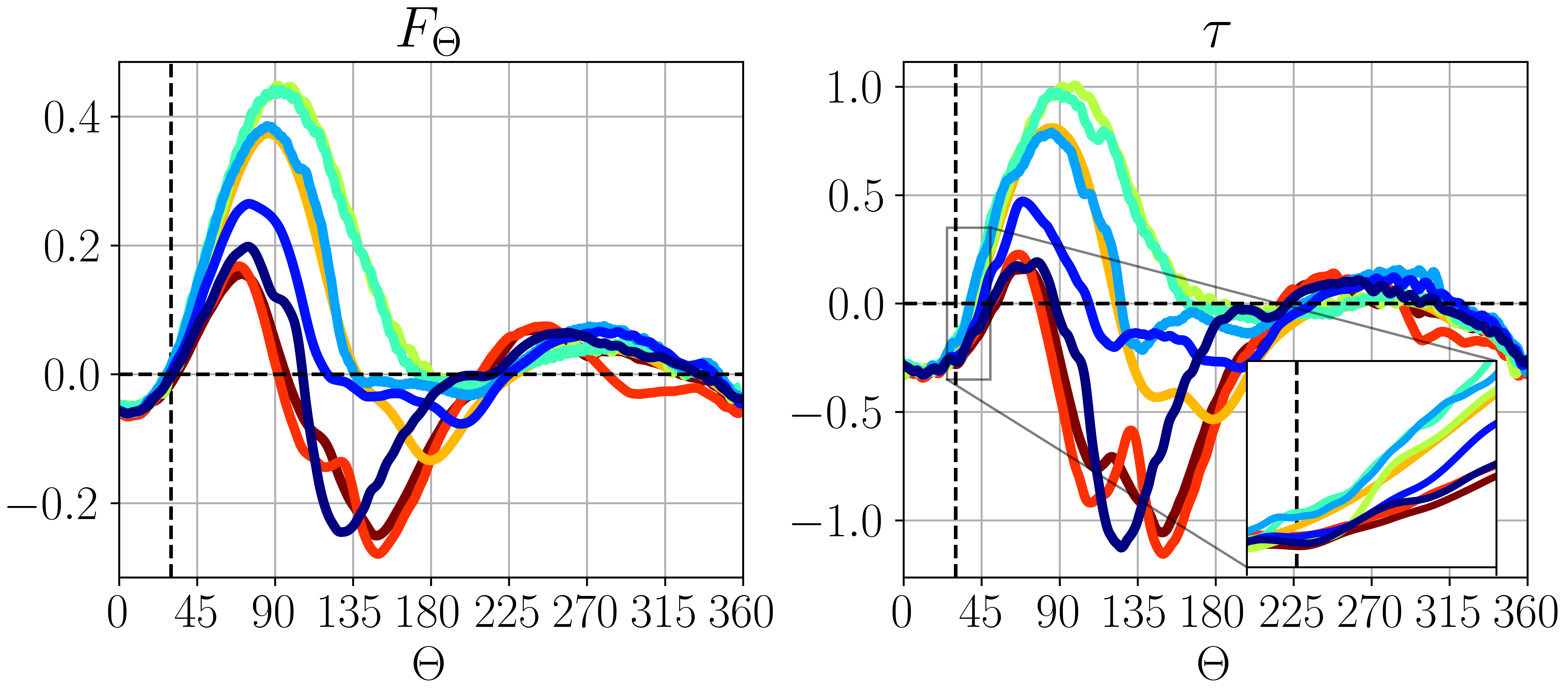}
        \put(60,3){a)}
        \put(560,3){b)}
    \end{overpic}
    \begin{overpic}[trim = 0cm 1cm 0cm 0cm,clip,width=.9\textwidth]{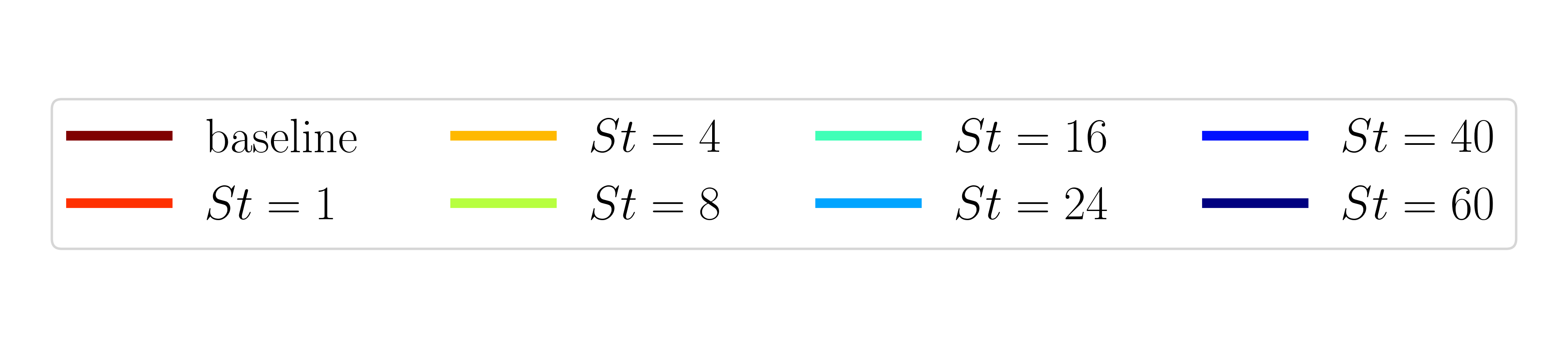}
    \end{overpic}    
\caption{Evolution of azimuthal force and torque for both baseline and controlled cases across different actuation frequencies: \(\text{St} = 1\), \(\text{St} = 4\), \(\text{St} = 8\), \(\text{St} = 16\),  \(\text{St} = 24\),  \(\text{St} = 40\), \(\text{St} = 60\). The panels display a) the azimuthal force coefficient, and b) the torque coefficient.}
\label{fig:control_coeffs} 
\end{figure}

Simulations are performed using different actuation frequencies within the range \(1 \leq \text{St} \leq 60\). Figure \ref{fig:control_coeffs}(a) presents the LES results in terms of the force in the azimuthal direction (\(F_\Theta\)) while the torque (\(\tau\)) is depicted in Fig. \ref{fig:control_coeffs}(b). The frame of reference for the $\Theta$ direction is illustrated in Fig. \ref{fig:motion}. The curves depict the evolution of each aerodynamic coefficient, phase averaged for each azimuthal position over 6 cycles. Different colors are used to represent the various actuation frequencies tested (\(\text{St} = 1\), \(\text{St} = 4\), \(\text{St} = 8\), \(\text{St} = 16\), \(\text{St} = 24\), \(\text{St} = 40\), \(\text{St} = 60\)) alongside the baseline case. A vertical black dashed line denotes the instant at $\Theta = 30^\circ$, when the flow control begins to induce a significant change in the trends of the aerodynamic coefficients. An inset is used in the torque plot for the purpose of visualization. This observation will subsequently be utilized to justify the selection of the time instant at which the resolvent analysis is conducted.

The curves reveal that actuation at a very low frequency (\(\text{St} = 1\)) or at very high frequencies ($\text{St} > 40$) fails to enhance the aerodynamic performance of the turbine. These results contrast with those reported by \citet{visbal_benton_control}, where dynamic stall at a Reynolds number O($10^5$) is initiated by leading-edge separation due to the bursting of a laminar separation bubble. In that case, the stall onset is governed by high-frequency dynamics of the bubble, indicating a fundamentally different mechanism. A comparison between the baseline case and those at $\text{St} = 1$, $40$ and $60$ shows that the overall trends in terms of $F_\Theta$ and torque $\tau$ remain similar. 
The azimuthal force indicates the presence of aerodynamic drag for \(\Theta > 90^\circ\) for the baseline and $\text{St} = 1$ cases, primarily caused by the formation of the DSV in the rear portion of the blade and the subsequent development of the TEV.
The baseline case shows a small window of useful torque production observed between \(45^\circ < \Theta < 75^\circ\) and the same observation is made for $\text{St} = 1$ and $60$. However, actuation at $\text{St} = 40$ slightly improves aerodynamic performance, producing useful torque in the range \(45^\circ < \Theta < 100^\circ\).

The results suggest that within the range of frequencies tested, aerodynamic performance improves at $4\leq \text{St} \leq 24$. Both the magnitude of the useful torque and the duration over which it is produced are significantly enhanced compared to the baseline case. However, it is evident that at \(\Theta \approx 120^\circ\), there is still a change in sign of \(\tau\) for $\text{St} = 4$ and $24$, indicating that the formation of the DSV may be delayed but not entirely eliminated, occurring at higher azimuthal angles.

Actuation at moderate frequencies \(\text{St} = 8\) and \(16\) provides the best aerodynamic performance of the turbine, extending the region of useful torque production to \(45^\circ \leq \Theta \leq 180^\circ\). Analysis of the azimuthal force distribution over the cycle reveals that this improvement is primarily due to the suppression of aerodynamic drag during this phase of the motion. In the absence of the drag induced by the DSV and TEV, the forces acting on the airfoil more closely resemble those of attached flow, aligning with the direction of motion and thereby improving efficiency. In terms of actuator efficiency, the relationship between \(W_{\text{jet}}\) and \(W_{\tau}\) is approximately \(1\%\) for $ 8 \leq \text{St} \leq 16$. This indicates that the work introduced by the jet is minimal compared to the work produced by the turbine, highlighting the efficiency of the actuation mechanism in improving the aerodynamic performance with relatively low energy input.

\begin{figure}[H]
    \centering

    \begin{overpic}[trim = 0cm 1cm 0cm 2cm,clip,width=0.99\textwidth]{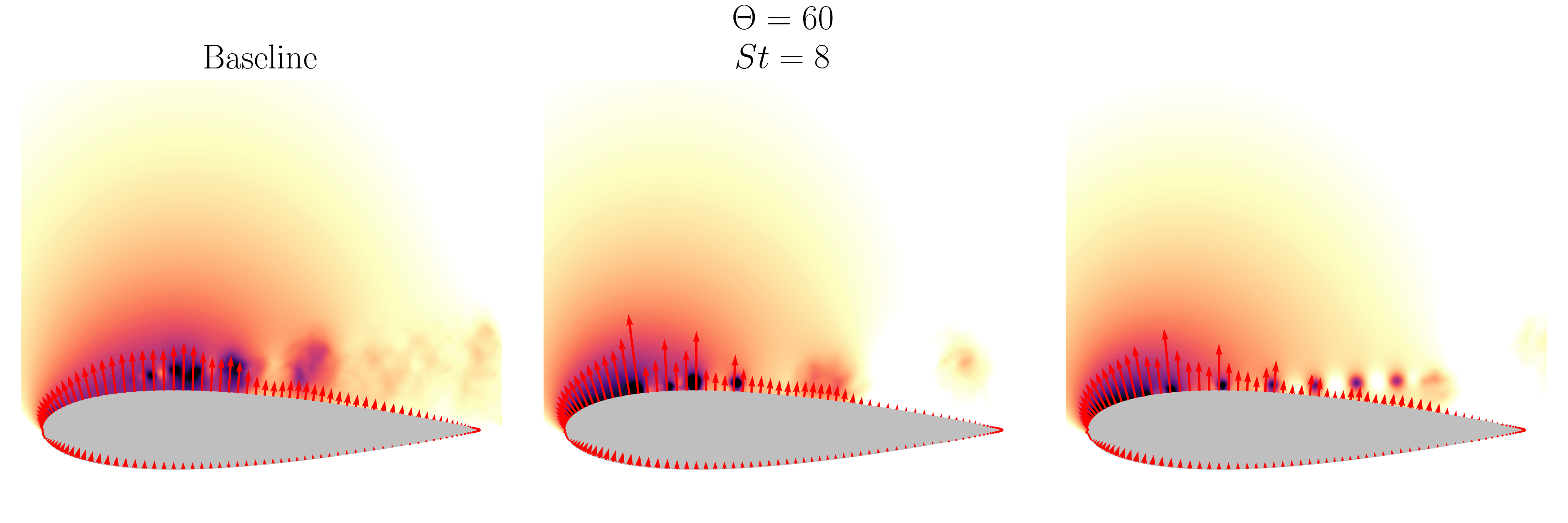}
        \put(30,170){$\Theta = 60$}
        \put(120,220){Baseline}
        \put(450,220){St = 8}
        \put(770,220){St = 24}
    \end{overpic}
    \begin{overpic}[trim = 0cm 1cm 0cm 2cm,clip,width=0.99\textwidth]{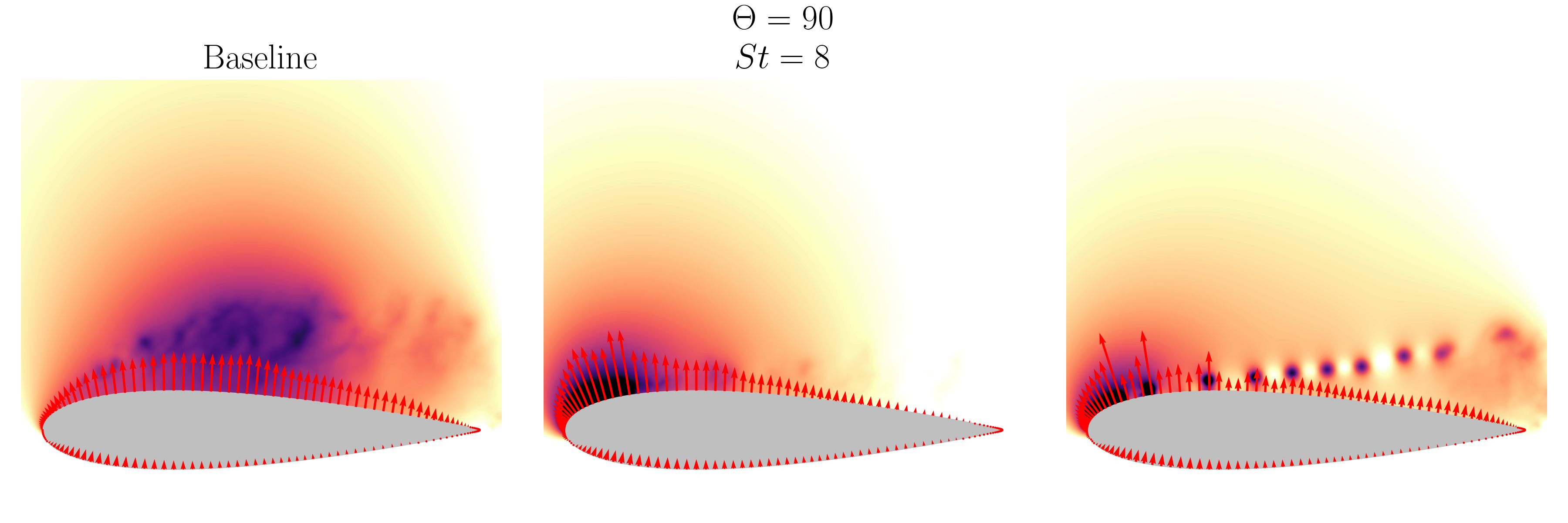}
        \put(30,170){$\Theta = 90$}
    \end{overpic}
    \begin{overpic}[trim = 0cm 1cm 0cm 2cm,clip,width=0.99\textwidth]{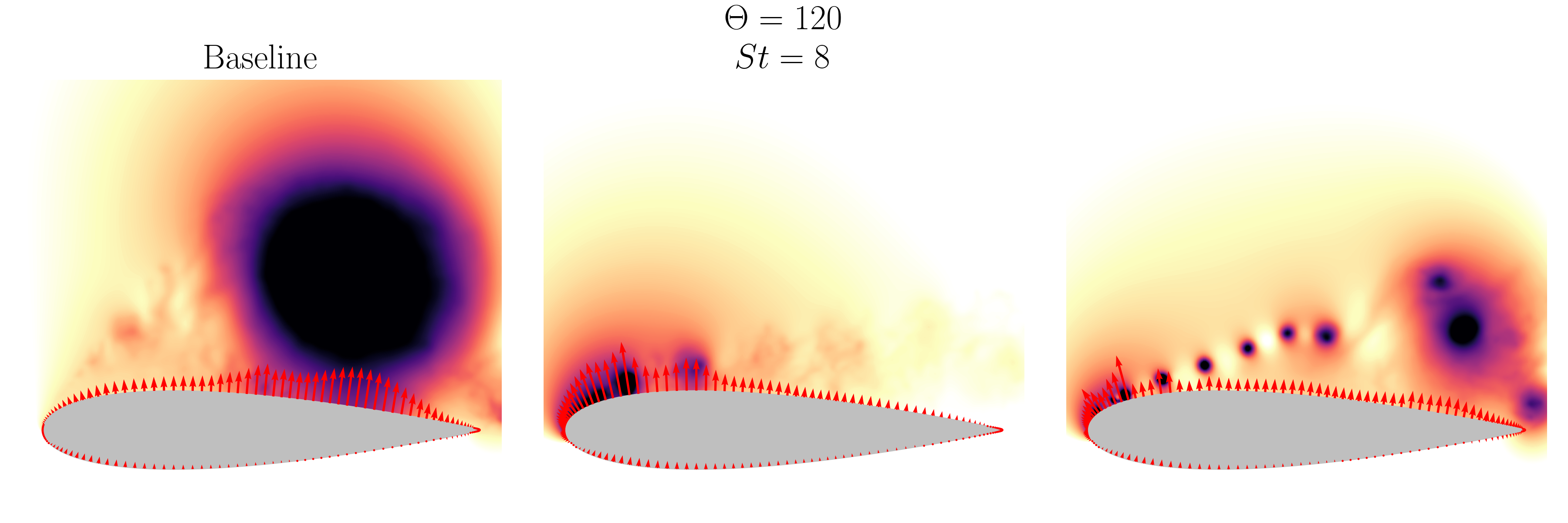}
        \put(30,170){$\Theta = 120$}
    \end{overpic}
    \begin{overpic}[trim = 0cm 1cm 0cm 2cm,clip,width=0.99\textwidth]{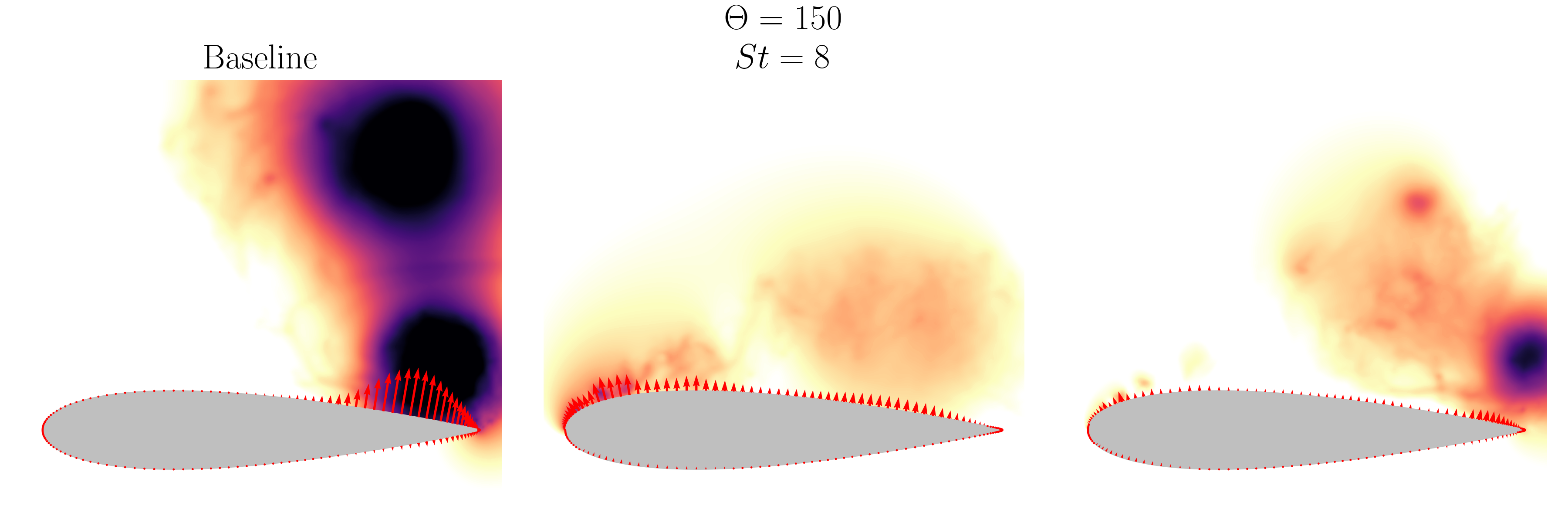}
        \put(30,170){$\Theta = 150$}
    \end{overpic}
    \begin{overpic}[trim = 0cm 0cm 0cm 0cm,clip,width=0.5\textwidth]{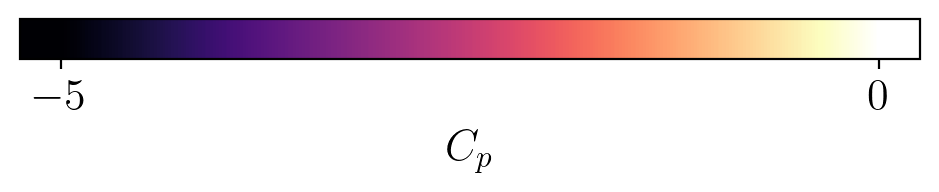}
    \end{overpic}
\caption{Evolution of instantaneous spanwise-averaged pressure coefficient (\(C_p\)) contours for three cases during selected azimuthal angles of the upwind motion. The left column depicts the flow fields for the baseline (uncontrolled) case, the center column presents the solutions of the actuated case with $\text{St} = 8$ while the right one showcases the actuated case with $\text{St} = 24$.}
\label{fig:control_contour} 
\end{figure}

\begin{figure}[H]
    \centering
    \begin{overpic}[trim = 0cm 0cm 0cm 0cm,clip,width=0.99\textwidth]{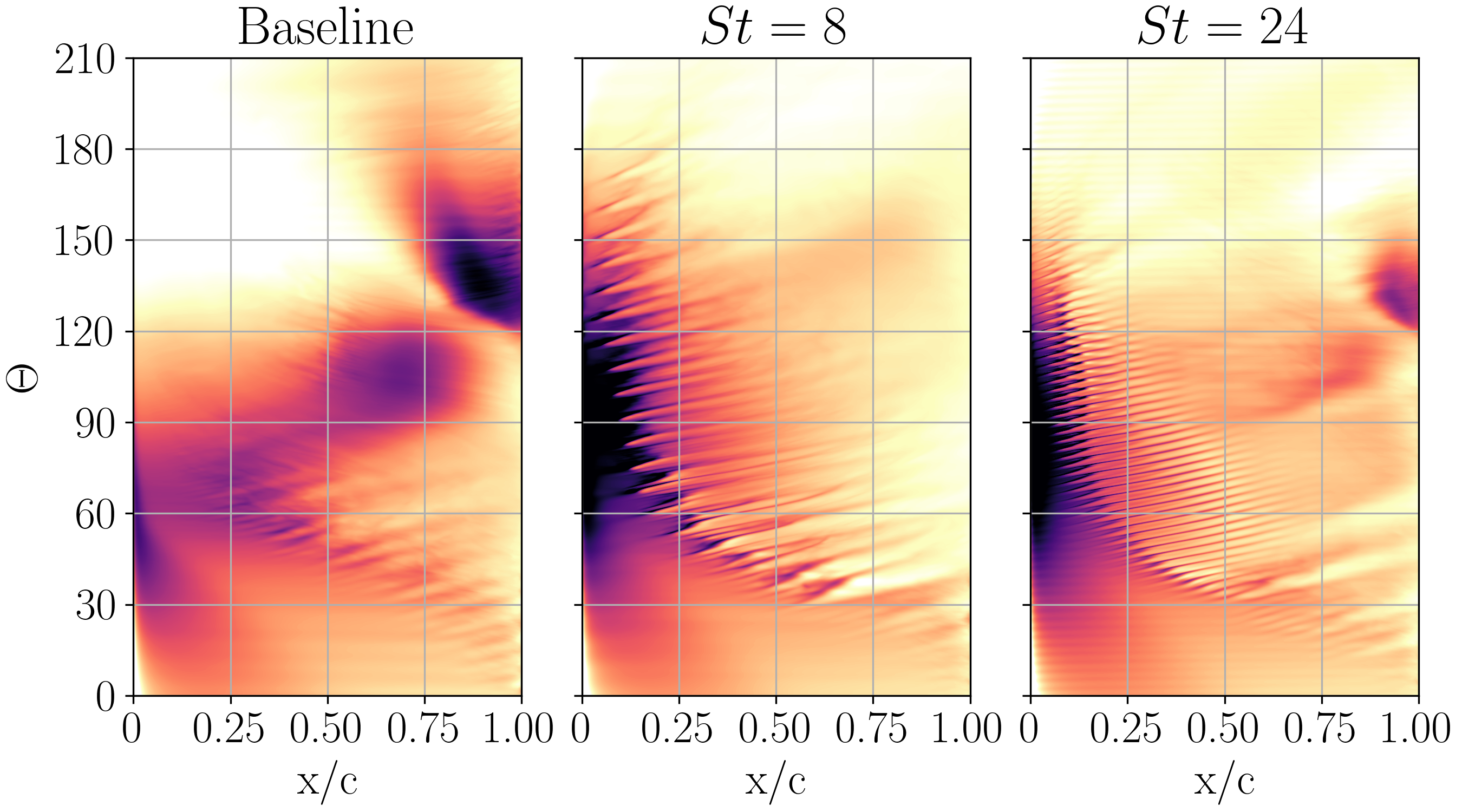}
    \end{overpic}
    \begin{overpic}[trim = 0cm 0cm 0cm 0cm,clip,width=0.5\textwidth]{cp_colorbar.png}
    \end{overpic}
    
\caption{Evolution of the spanwise-averaged surface pressure coefficient (\(C_p\)) on the blade suction side for the baseline and actuated cases. The left column depicts the results for the baseline (uncontrolled) case, the center column presents the solution of the actuated case with $\text{St} = 8$ while the right one showcases the actuated case with $\text{St} = 24$.}
\label{fig:cp_surf} 
\end{figure}

To better understand the changes in flow topology and separation behavior induced by the control strategy, Figs. \ref{fig:control_contour} and \ref{fig:cp_surf} present spanwise-averaged pressure coefficient (\(C_p\)) contours for the flow fields near the turbine and along the airfoil surface at selected azimuthal positions, respectively. In figure \ref{fig:control_contour} the distribution of forces on the surface of the airfoil is also shown at each instant to better elucidate the thrust generation mechanism provided by flow control. For the sake of brevity only results of the baseline, $\text{St} = 8$ and $\text{St} = 24$ cases are shown. At \(\Theta = 60^\circ\), Kelvin-Helmholtz instabilities are observed along the shear layer on the suction side, forming near the leading edge, where a low-pressure region is observed. Different wavelengths of the instabilities are observed in Fig. \ref{fig:cp_surf} for the uncontrolled and controlled cases because of the different actuation frequencies. In the controlled cases, more organized disturbances are induced by actuation and their spatial scales reduce with increasing frequencies. Moreover, they form further upstream as the shear layer develops toward the leading edge. As shown in Fig. \ref{fig:cp_surf}, the leading edge suction peaks (negative regions of $C_p$) of the controlled cases are also notably stronger compared to that of the baseline. 

Figures \ref{fig:control_contour} and \ref{fig:cp_surf} show that, in the interval \(60^\circ < \Theta < 90^\circ\), the baseline case exhibits a significant reduction in suction on the leading edge, leading to flow separation and the formation of a coherent vortex in the region \(0.25 < x/c < 0.5\). In contrast, the actuated cases maintain attached flows, generating a sustained suction region over \(0 < x/c < 0.25\). Subsequently, for \(90^\circ < \Theta < 120^\circ\), the shear layer formed at the leading edge feeds the growth of the DSV in the baseline case. However, for the actuated cases, the flow remains attached near the leading edge, but with discrete vortices being ejected. These smaller-scale vortices do not coalesce into a coherent DSV. During this phase of the cycle, the surface pressure contours show that both controlled cases preserve a pronounced suction peak near the leading edge, while in the baseline case, the low-pressure region that indicates the trace of the DSV is transported toward the aft portion of the airfoil, promoting aerodynamic drag.

In the actuated cases, the leading-edge suction persists up to \(120^\circ < \Theta < 150^\circ\), as shown in Fig. \ref{fig:cp_surf}. This figure also shows the trace of the TEV on the airfoil surface for the baseline case for the same phase of the cycle. As presented in Fig. \ref{fig:control_contour}, both the DSV seen at $\Theta = 120^\circ$ and the TEV shown at $\Theta = 150^\circ$ lead to lower pressure regions over the airfoil aft surface, which in turn lead to a drag increase. 
On the other hand, the actuated case at $\text{St} = 8$ depicts a strong suction peak on the leading edge and no sign of the DSV and TEV. These features lead to a higher circulation over the airfoil, and an increase in torque during a pronounced period of motion. For $\text{St} = 24$, the discrete vortices ejected from the leading edge coalesce on the trailing edge, leading to the formation of a small TEV that is responsible for reducing the circulation over the airfoil, also causing a lower thrust for this particular case.

\subsection{Insights of flow actuation from linear stability and resolvent analyses}

The physical mechanisms underlying the effectiveness of the proposed flow control strategy are investigated in this section through the application of bi-global stability and resolvent analyses. These techniques improve the understanding of vortex dynamics in the separated shear layer and wake regions of the blade. This analysis provides a more comprehensive explanation of why a particular actuation frequency is capable of promoting greater improvements in aerodynamic performance.

The present linear analysis is performed for a particular phase of the cycle. Here, the selected time instant for this investigation corresponds to the notable change in aerodynamic behavior, as indicated by the control results presented in Fig.~\ref{fig:control_coeffs}. It is observed that disturbances introduced by the jet actuation begin to significantly modify the azimuthal force and torque at $\Theta = 30^\circ$. Consequently, this specific phase of the blade motion is chosen for the subsequent analysis. The investigation is conducted in a non-inertial reference frame centered at the quarter-chord position of the blade. All flow variables are non-dimensionalized using the corresponding freestream reference quantities. It is also worth emphasizing that we assume the time evolution of the base flow to be negligible relative to the timescale of the instabilities being analyzed. Furthermore, we assume that the effects of centrifugal and Coriolis forces are negligible \cite{Gardner_2019}.

\begin{figure}[H]
    \centering
    \begin{overpic}[trim = 0cm 0cm 0cm 0cm,clip,width=0.999\textwidth]{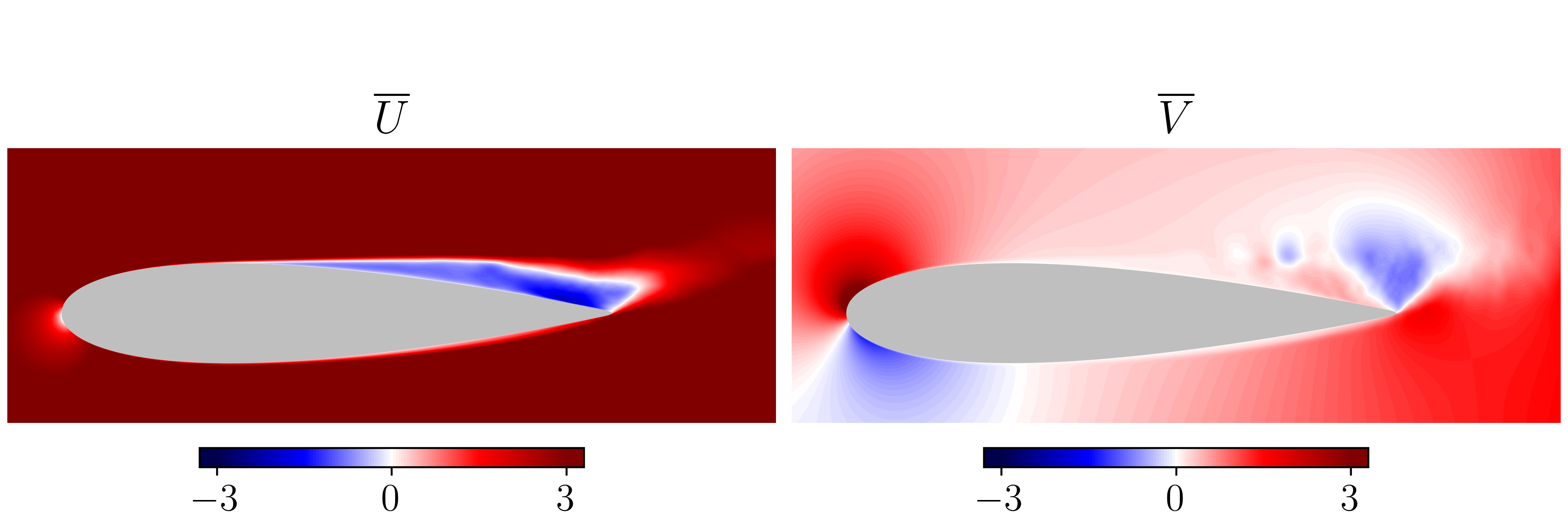}
    \end{overpic}
\caption{Spanwise- and phase-averaged $\overline{U}$ and $\overline{V}$ velocity fields at $\Theta = 30^\circ$ employed as base flows in the linear operator.}
\label{fig:baseflow} 
\end{figure}

The base flow utilized in both the stability and resolvent analyses is obtained by first performing a spanwise averaging, followed by a phase averaging of the LES data over 6 cycles of the blade motion. Due to the spanwise homogeneity, the mean spanwise velocity component is null, i.e., $\overline{W} = 0$. Figure~\ref{fig:baseflow} presents the contours of the mean velocity components in the x and y directions, denoted by $\overline{U}$ and $\overline{V}$, respectively. It is important to emphasize that, since a compressible formulation is employed, the mean density ($\overline{\rho}$) and temperature ($\overline{T}$) fields are also considered in the analysis. Furthermore, since the solution is obtained through phase averaging, there is no guarantee that all unsteady instabilities will be filtered out, as would be the case with a time-averaged solution around a stationary airfoil. As a result, flow structures that are recurrent during a particular phase of the motion will be present in the base flow, as can be seen in the aft portion of the airfoil. Despite this limitation, we ensure the convergence of the eigenvalue spectrum with respect to the phase-averaging process, by considering 5, 6 and 7 cycles. Here, the results obtained by considering 6 cycles are reported. This verification process is particularly important for high-frequency modes, which are more challenging to resolve because of their association with short-wavelength structures that are highly sensitive to small local variations in the base flow.

At the instant considered for the analysis ($\Theta = 30$), the blade is at the beginning of the upwind motion, near the phase corresponding to the maximum relative velocity ($\Theta = 0$). A prominent region of flow reversal is observed in the aft portion of the airfoil for the mean streamwise velocity component $\overline{U}$. This feature is highlighted by the white color in the contour that delimits the $\overline{U}=0$ region. This reversed flow region is accompanied by the development of a shear layer originating near the quarter-chord position and extending toward the trailing edge. As previously discussed, Kelvin-Helmholtz instabilities arising within the shear layer are responsible for the accumulation of negative vorticity, which ultimately feeds the formation of the DSV. This, in turn, results in a substantial increase in aerodynamic drag, thereby hindering the efficiency of energy extraction.

\begin{figure}[H]
    \centering
    \begin{overpic}[trim = 0cm 0cm 0cm 0cm,clip,width=\textwidth]{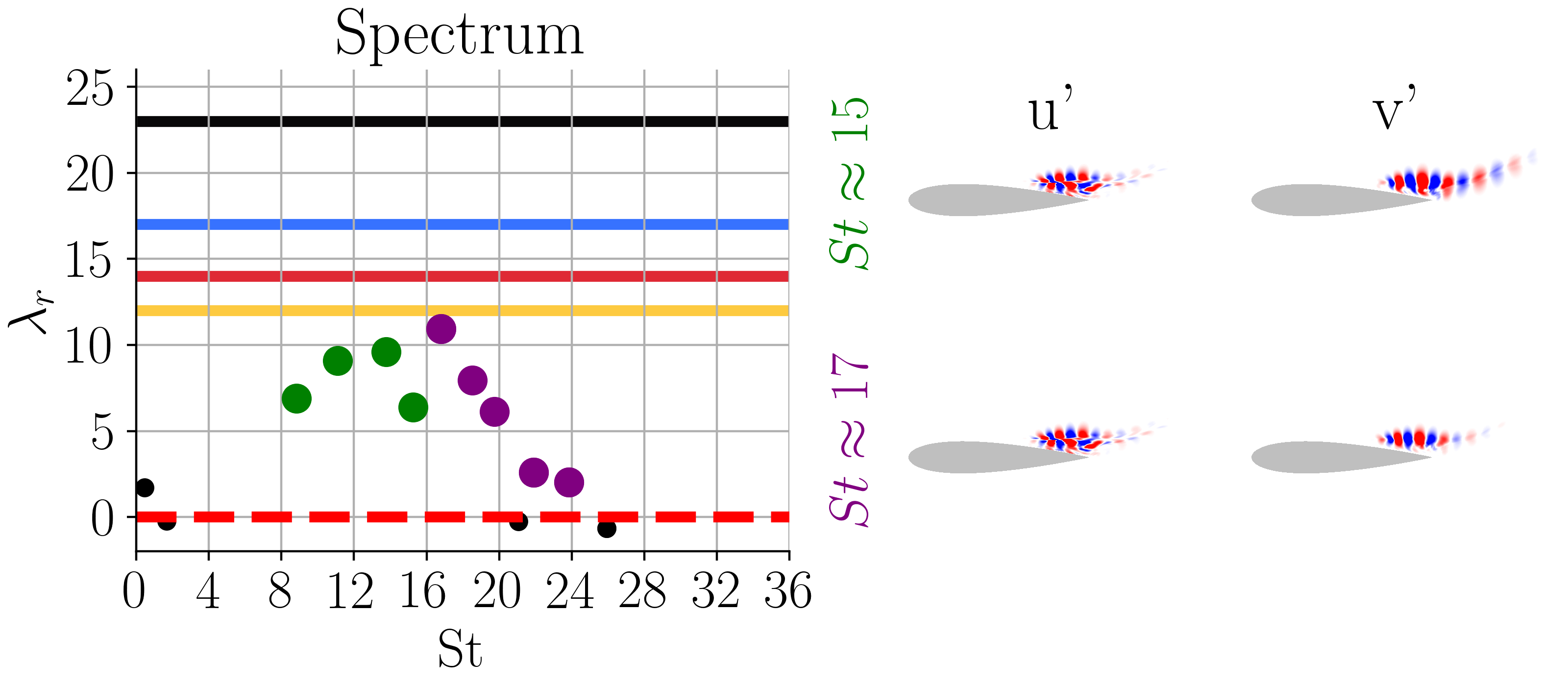}
    \end{overpic}
    \begin{overpic}[trim = 0cm 0cm 0cm 0cm,clip,width=\textwidth]{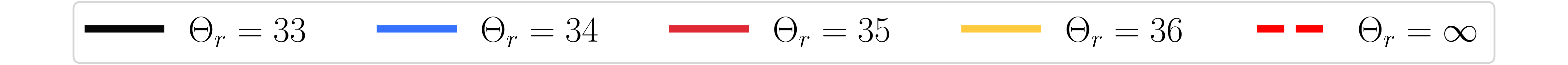}
    \end{overpic}
\caption{Eigenvalue spectrum computed for $\Theta = 30^\circ$ (left). Unstable modes that portray a coupling mechanism between the shear layer and the wake are shown by green dots, while those associated exclusively with the shear layer are shown by purple dots. The red dashed line delimits the region of instability and solid colored lines represent the discounting parameter that leads to an analysis of the system at future azimuthal positions $\Theta_r$. Spatial support of the eigenmodes computed for $\Theta = 30^\circ$ near the frequency where the dynamics changes ($\text{St} \approx 16$) from wake-shear-layer coupled modes to shear-layer-dominated modes (right).
}
\label{fig:spectrum} 
\end{figure}


The results of the linear stability analysis of the base flow are presented in Fig. \ref{fig:spectrum}. The eigenvalue spectrum, shown in terms of Strouhal number and the real part of the eigenvalues (growth rates), reveals that the flow is linearly unstable over a broad frequency range ($0 \leq \text{St} \leq 24$). A dashed red line marks the boundary of the instability region, with eigenvalues represented by circular markers. Modes located above this line are unstable. Among these, green markers highlight eigenmodes with spatial support in both the shear layer and the wake, indicating a coupling mechanism between these flow features. Purple markers, in contrast, represent modes primarily confined to the shear layer, characterizing Kelvin--Helmholtz instabilities. The plots on the right illustrate this transition in modal behavior, displaying velocity eigenmodes representative of both categories near $\text{St} = 16$, which appears to act as an upper boundary threshold frequency separating wake-shear-layer coupled dynamics from shear-layer-dominated behavior. This distinction is evident in the extent of the spatial support of the $v'$ modes toward the wake, which becomes more pronounced at lower frequencies.



The result obtained by the present bi-global linear stability analysis align with findings from \citet{yeh_taira_2019}. Their work showed that, for the static stall over an airfoil, low-frequency modes are associated with the wake dynamics. On the other hand, moderate frequency modes display a coupling between the shear layer and wake, whereas  high frequencies are exclusively related to Kelvin-Helmholtz instabilities in the shear layer. The spatial support of these modes also agrees with the present LES results, in which vorticity accumulation near the trailing edge is attributed to the growth of Kelvin-Helmholtz structures. Furthermore, the wavelengths of the stability modes align well with those observed in the LES, validating the linear framework. It is also important to note that the timescale associated with the most unstable mode is approximately 90 times faster than that of the VAWT motion, characterized by $\text{St} \approx 0.19$. Therefore, for the purposes of the finite-time analysis to be conducted, the temporal evolution of the base flow can be neglected.


The present linear stability analysis reflects the natural response of the system as described in equation \ref{eq:LSA}. However, for flow control purposes, the response of the forced system given by equation \ref{eq:final_resolvent} is more relevant. In this context, applying an actuation at a frequency near a given eigenvalue can trigger a resonance effect, effectively steering the system response toward the associated eigenvector. The identification of appropriate bi-global modes is therefore crucial in designing effective actuation strategies. Modes that couple the shear layer and the wake are particularly desirable, as they enhance the transport of vorticity downstream and help preventing the DSV formation, which is attributed to vorticity accumulation. A similar approach was found by \citet{Nair_Yeh_Kaiser_Noack_Brunton_Taira_2019}, where the optimal control input for drag reduction in a airfoil under static stall was shown to excite both the shear layer frequency as well as the harmonics associated with the wake dynamics.


To examine the forced response of the system, the discounted resolvent analysis framework is employed. According to \citet{Yeh_taira_2020}, the most appropriate values for analyzing the finite-time system response should ensure that all eigenvalues of the discounted system are stable, that is, $\psi \geq \lambda_r = 11$ (growth rate of the most unstable eigenvalue). Hence, different discounting values are considered: $\psi = 23$, $\psi = 17$, $\psi = 14$, and $\psi = 12$. Since these parameters are related to the finite time window of the analysis, they can also be represented in terms of the blade angular displacement as $\Delta \Theta(\psi) =180\omega /\psi  \pi$ and the final azimuthal position of the finite window can be computed as $\Theta_r = 30^\circ + \Delta \Theta(\psi)$. In the analysis that follows, we use the latter equation to express the results in terms of the respective finite-time azimuthal positions: $\Theta_r(\psi = 23) = 33^\circ$, $\Theta_r(\psi=17) = 34^\circ$, $\Theta_r(\psi = 14) = 35^\circ$,  $\Theta_r(\psi = 12) = 36^\circ$. 

\begin{figure}[H]
    \centering
    \begin{overpic}[trim = 0cm 0cm 0cm 0cm,clip,width=0.89\textwidth]{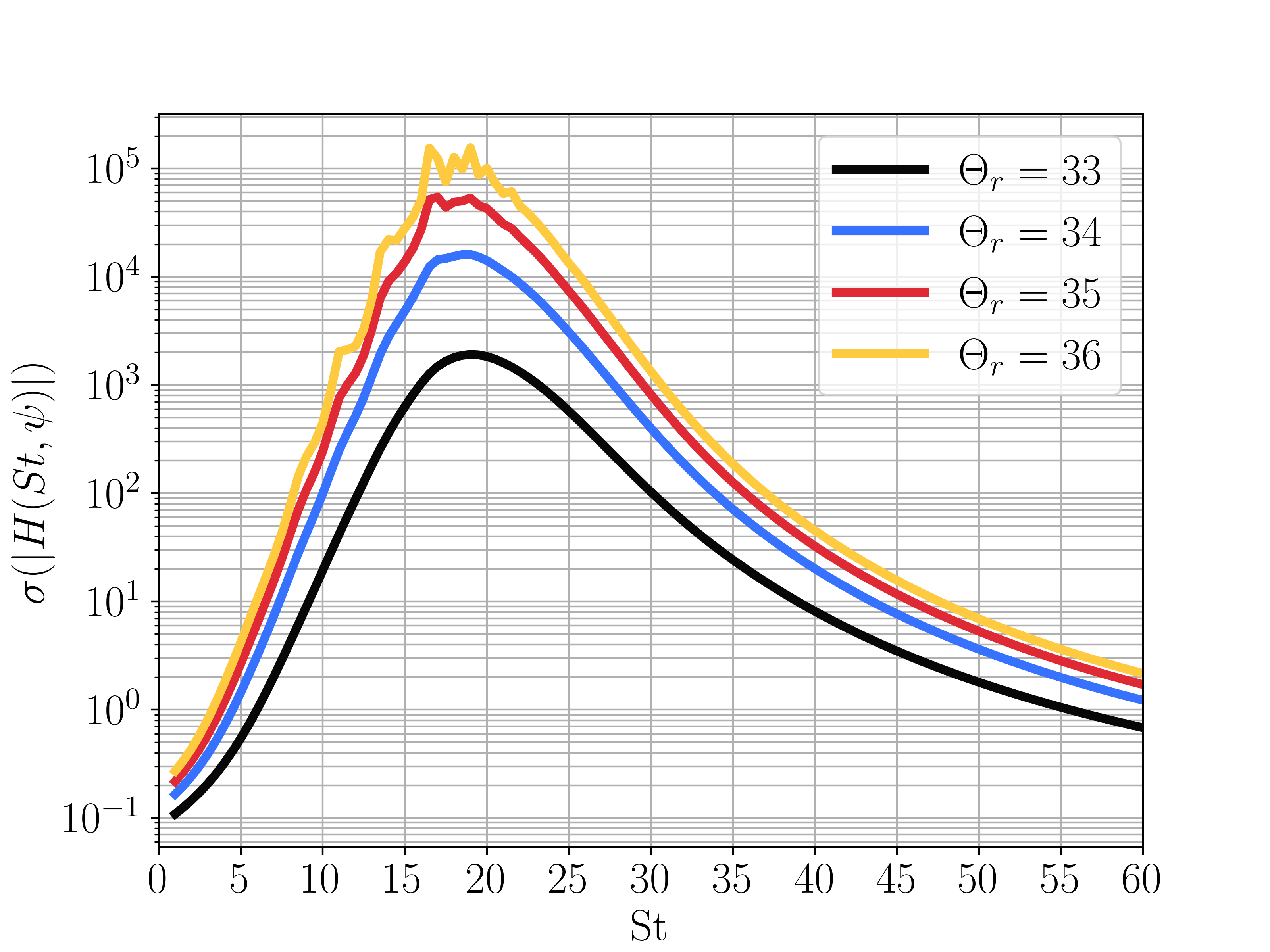}
    \end{overpic}
\caption{Resolvent gains for the airfoil at $\Theta = 30^\circ$ considering different discounting parameters (finite-time windows).}
\label{fig:resolvent} 
\end{figure}

The curves with the resolvent gains are depicted in Fig.~\ref{fig:resolvent} for the different discounting parameters discussed previously. The corresponding color-coded paths are indicated in the eigenvalue spectrum shown in Fig.~\ref{fig:spectrum}. The curves reveal that the largest amplification of the system over a finite time interval, for an observation window up to $\Theta_r(\psi = 12) = 36^\circ$, occurs at $\text{St} \approx 16$. The gain profiles reveal a significant amplification of disturbances in the range $4 \leq \text{St} \leq 50$, for this finite-time window. This result is consistent with the control improvements observed in the LES simulations, where actuation at similar frequencies promoted flow reattachment and enhanced aerodynamic performance. As the value of $\psi$ decreases, the gain profiles change due to increased proximity to the most unstable eigenvalues. This effect is particularly evident when comparing the curves for $\Theta_r(\psi = 23) = 33^\circ$ and $\Theta_r(\psi = 12) = 36^\circ$. As the discounting parameter approaches the growth rate of the most unstable eigenvalue, $\lambda_r = 11$, a peak in the resolvent gain clearly emerges at the frequency associated with this eigenvalue. Similar behavior was also reported by \citet{yeh_taira_2019} and \citet{Yeh_taira_2020}.

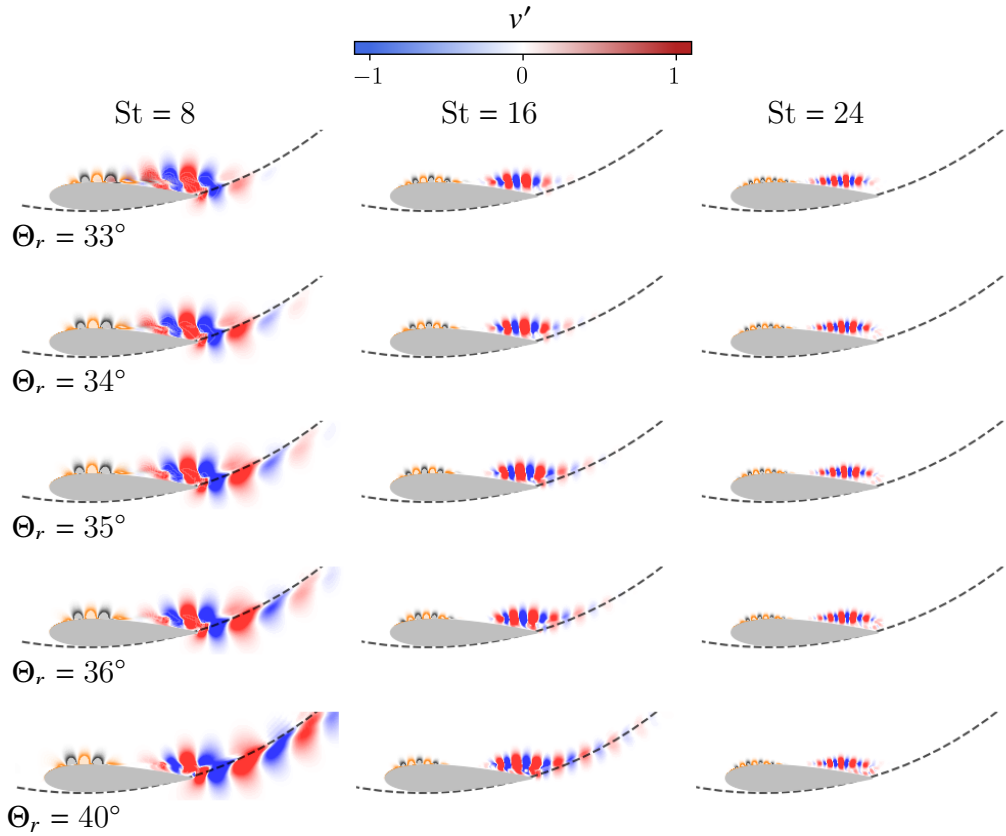
\begin{figure}[H]
    \centering
    \begin{overpic}[trim = 0cm 0cm 0cm 0cm,clip,width=0.35\textwidth]{lcs_colorbar.png}
    \put(460,200){$v'$}
    \end{overpic}
    \begin{overpic}[trim = 0cm 0cm 0cm 0cm,clip,width=0.99\textwidth]{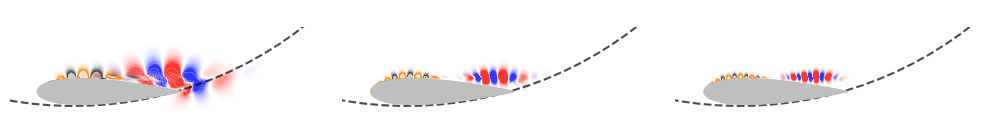}
    \put(0,0){$\Theta_r = 33^\circ$}
    \put(100,120){$\text{St} = 8$}
    \put(420,120){$\text{St} = 16$}
    \put(740,120){$\text{St} = 24$}
    \end{overpic}
    \begin{overpic}[trim = 0cm 0cm 0cm 0cm,clip,width=0.99\textwidth]{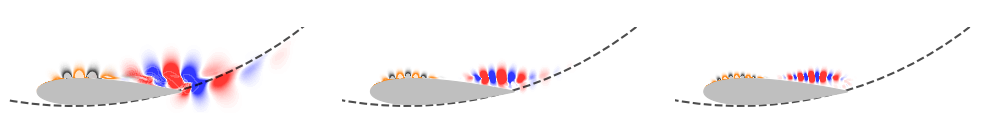}
    \put(0,0){$\Theta_r = 34^\circ$}
    \end{overpic}
    \begin{overpic}[trim = 0cm 0cm 0cm 0cm,clip,width=0.99\textwidth]{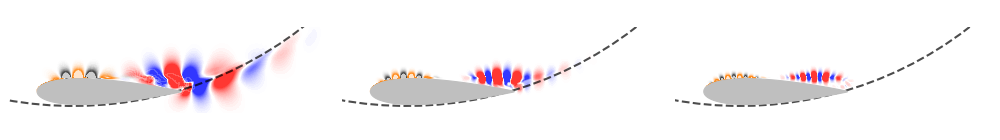}
    \put(0,0){$\Theta_r = 35^\circ$}
    \end{overpic}
    \begin{overpic}[trim = 0cm 0cm 0cm 0cm,clip,width=0.99\textwidth]{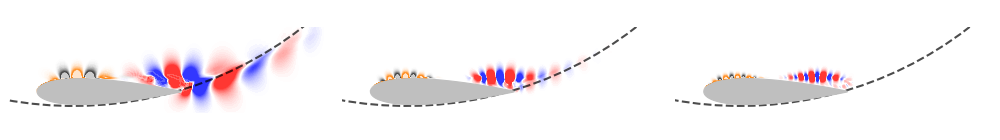}
    \put(0,0){$\Theta_r = 36^\circ$}
    \end{overpic}
    \begin{overpic}[trim = 0cm 0cm 0cm 0cm,clip,width=0.99\textwidth]{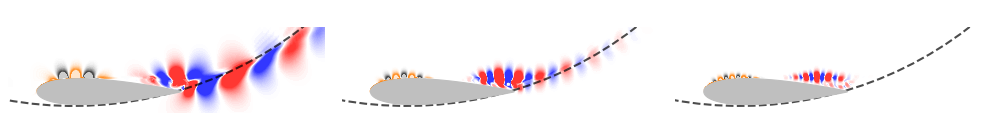}
    \put(0,0){$\Theta_r = 40^\circ$}
    \end{overpic}
\caption{Temporal evolution of the discounted resolvent forcing (gray-orange) and response (blue-red) modes for different actuation frequencies $\text{St} = 8$, $16$ and $24$. The dashed line highlights the blade trajectory.}
\label{fig:resolvent_evo} 
\end{figure}

To gain a more comprehensive understanding of how disturbances evolve across different timescales and actuation frequencies, and how they promote flow reattachment and aerodynamic improvement, we analyze the resolvent response modes for the four previously defined time windows. An additional case at $\Theta_r(\psi = 7) = 40^\circ$ is also shown to emphasize the convective interpretation of the discounted resolvent analysis. Figure \ref{fig:resolvent_evo} presents a comparison of the forcing and response modes for three representative actuation frequencies, $\text{St} = 8$, $16$ and $24$. We selected these particular frequencies to highlight how the coupled response of the wake and shear layer changes with increasing actuation frequency. According to the spectrum, low frequencies are associated with modes that involve both the shear layer and the wake, whereas for $\text{St} \gtrapprox 16$, the response shifts toward shear-layer-dominated modes. The visualizations are arranged vertically to illustrate the temporal evolution of each mode from top to bottom. The modes are visualized in a non-inertial frame of reference, with the theoretical trajectory of the blade trailing edge overlaid as black dashed lines to highlight the airfoil wake. In each contour plot, the spatial support of the forcing mode is shown in gray and orange, while the response modes are depicted in blue and red. The left column displays the evolution of the modes for the forcing frequency of $\text{St} = 8$, the center column shows the modes obtained for $\text{St} = 16$, and the right column presents the corresponding evolution for the higher actuation frequency at $\text{St} = 24$. It is important to notice that, for all cases, the forcing modes display a spatial support on the suction side, toward the leading edge. This is an important result that justifies the actuation upstream of this region. In this case, the disturbances would always excite the optimal forcing regions, leading to their respective optimal responses.

An analysis of the resolvent modes reveals that, for all frequencies analyzed, the initial optimal forcing and system response are predominantly concentrated within the shear layer, giving rise to Kelvin-Helmholtz instabilities that locally promote flow reattachment as they are advected downstream. Over time, the spatial support of the response modes tends to shift toward the trailing edge, while the forcing modes remain localized near the separation point. However, when comparing the response modes across different frequencies, it becomes evident that the spatial structures of the lower-frequency modes extend into the wake, whereas higher-frequency modes remain confined to the trailing edge, regardless of the time window considered. This behavior suggests a limited ability of high-frequency, short-wavelength vortices to propagate into the wake region. This mechanism appears to play a key role under the current flow conditions, where the dynamic stall regime is characterized by trailing-edge stall, with vorticity accumulation initiating from the trailing edge.

In the present large-eddy simulations, these high-frequency disturbances undergo vortex pairing, which leads to vorticity accumulation at higher azimuthal positions of the VAWT cycle. This accumulation results in the formation of a coherent vortex near the trailing edge, as shown in the last column of Fig. \ref{fig:control_contour} for $\Theta = 120^\circ$. This flow feature causes a reduction in aerodynamic performance compared to actuation at moderate frequencies ($8 \leq \text{St} \leq 16$). In contrast, the spatial structure of the response for an excitation at $\text{St} = 8$, which is closer to the natural coupling frequency between the shear layer and the wake, demonstrates that the resulting vortices are effectively transported downstream. This flow feature mitigates vorticity buildup near the trailing edge and maintains flow attachment, improving aerodynamic performance.

The transport behavior inferred from the linear analysis is further validated by observations from the nonlinear simulations. Although the linear analysis is performed for a fixed base flow at $\Theta = 30^\circ$, the results suggest that enhanced downstream advection of disturbances associated with intermediate frequency actuation persists over a broader phase range. The LES results confirm this effect over different phases of the motion, for instance in Fig.\ref{fig:cycles}(b), which shows a well-organized Von Kármán vortex street in the controlled case with $\text{St} = 8$ at $\Theta = 0^\circ$. During the onset of dynamic stall, this effect becomes more prominent in the interval $70^\circ \leq \Theta \leq 90^\circ$. Therefore, actuation at an intermediate frequency not only suppresses local flow separation but also enables the generated structures to propagate efficiently into the wake. This reduces the amount of accumulated vorticity near the airfoil surface, hindering the formation of large-scale coherent vortices. This, in turn, increases the circulation over the airfoil, minimizing drag and enhancing the overall turbine performance. A movie showing the flow evolution computed by the LES is provided as supplementary material to support this interpretation, which is also consistent with the linear analysis.

\section{Conclusions}
\label{sec4}

This work investigates the aerodynamic performance and flow control of a NACA0018 airfoil operating as a single-bladed VAWT using wall-resolved LES. The high-fidelity simulations successfully capture key features of the dynamic stall phenomenon, demonstrating good agreement with trends in the literature. 
The detrimental effects of dynamic stall on the turbine performance are characterized, with drag forces arising from the DSV and TEV that limit energy production. 
Active flow control using suction and blowing actuation is demonstrated as an effective approach to mitigate these effects. Disturbances introduced in the boundary layer at appropriate actuation frequencies delay flow separation, maintain a suction peak at the leading edge, and suppress the formation of large coherent vortices, particularly at moderate actuation frequencies. 

Bi-global stability and resolvent analyses are employed to understand the effectiveness of a particular frequency range. Resolvent analysis performed with discounting, i.e., for a finite time window, shows that the excitation of coupled response modes on the shear layer and wake are more effective in advecting coherent structures. Hence, the Kelvin-Helmholtz instabilities excited by the actuation promote the local flow reattachment and are efficiently transported to the wake, without contributing to vorticity accumulation that leads to a coherent DSV. These improvements extend the portion of the cycle where useful torque is produced by the VAWT, significantly enhancing its aerodynamic performance.
%
%

The present actuation approach is highly energy-efficient and robust to variations in the incoming wind direction. The energy input from the suction and blowing jet accounts for only 1\% of the work produced by the turbine. The analysis of pressure coefficient contours highlights differences in flow behavior for controlled and baseline (uncontrolled) cases, showing the suppression of flow separation and a sustained suction peak near the leading edge for controlled scenarios.

In summary, this study demonstrates the potential of active flow control to enhance the performance of VAWTs, offering an efficient solution to mitigate dynamic stall and improve energy production. The numerical framework presented and its findings provide valuable insights for the design and optimization of next-generation wind energy systems.

\section*{Funding Sources}

The authors acknowledge Fundação de Amparo à Pesquisa do Estado de São Paulo, FAPESP, for supporting the present work under research grants No. 2013/08293-7, 2019/17874-0, 2021/06448-0, 2022/09196-4, 2022/08567-9, and 2024/20547-9. Conselho Nacional de Desenvolvimento Científico e Tecnológico, CNPq, is also acknowledged for supporting this research under grant No.\ 304320/2024-2.

\section*{Acknowledgments}

CEPID-CCES and CENAPAD-SP are acknowledged for providing the computational resources for this research through clusters Coaraci and Lovelace, respectively.

\appendix
\section{Governing equations}
\label{appendix}

The compressible Navier-Stokes equations are solved in non-dimensional form. 
The characteristic length, velocity components, density, pressure, and temperature are given, respectively, by the airfoil chord $c$, freestream speed of sound $a_{\infty}$, freestream density $\rho_{\infty}$, $\rho_{\infty}a_{\infty}^{2}$, and $(\gamma-1)T_{\infty}$. Here, $T_{\infty}$ is the freestream temperature and $\gamma$ is the ratio of specific heats. 
In a non-inertial system attached to the airfoil, the continuity, momentum, and energy equations are written as
\begin{equation}
\frac{\partial{\rho}}{\partial t} + \mbox{div} \, (\rho \boldsymbol{u}') = 0 \mbox{ ,}
\label{eq:continuity}
\end{equation}
\begin{eqnarray}
\frac{\partial (\rho \boldsymbol{u}')}{\partial t} 
+ \mbox{div} \, (\rho \boldsymbol{u}' \boldsymbol{u}' + p \boldsymbol{\mathrm{I}} - \boldsymbol{\tau}')
= 
- 2 \rho \boldsymbol{\Omega} \times \boldsymbol{u}' -
\rho \boldsymbol{\Omega} \times ( \boldsymbol{\Omega} \times \boldsymbol{x}' )
\mbox{ ,}
\label{eq:momentum}
\end{eqnarray}
and
\begin{equation}
\frac{\partial E}{\partial t} 
+ \mbox{div} \, \left[(E + p) \boldsymbol{u}' - \boldsymbol{\tau}' \cdot \boldsymbol{u}' + \boldsymbol{q} \right] = 0
\mbox{ ,}
\label{eq:energy}
\end{equation}
where $\boldsymbol{\Omega} = \left(0,0,\omega^* \right)$ is the angular velocity of the non-inertial frame and $\omega^{*}$ is the non-dimensional angular velocity. Primed quantities represent variables measured with respect to the moving frame. 

The total energy and heat flux for a fluid obeying Fourier law read, respectively, as
\begin{equation}
E = \frac{p}{\gamma - 1} + \frac{1}{2} \, \rho \left[ \boldsymbol{u}' \cdot \boldsymbol{u}' + \left(\boldsymbol{\Omega} \times \boldsymbol{x}' \right) \cdot  \left( \boldsymbol{\Omega} \times \boldsymbol{x}' \right) \right] \mbox{ ,}
\label{eq:total_energy}
\end{equation}
and
\begin{equation}
\boldsymbol{q} = -\frac{\mu M}{\mbox{Re}_\omega \mbox{Pr}} \, \mbox{grad} \, T \mbox{ .}
\end{equation}
Since the Navier-Stokes equations are written in a non-inertial frame of reference, the motion effects are accounted for by reevaluating the right-hand side of the momentum equation \ref{eq:momentum} and the total energy equation \ref{eq:total_energy} at every timestep. This approach entails no overhead, as the mesh can be kept fixed for any displacements of the airfoil.
Finally, the viscous stress tensor is given by
\begin{eqnarray}
\boldsymbol{\tau}'= \frac{M}{\mbox{Re}_\omega} \left[ 2 \mu \boldsymbol{S}' - \frac{2}{3} \mu \, \mbox{div} \, \boldsymbol{u}' \, \boldsymbol{\mathrm{I}} \right]
\mbox{ ,}
\end{eqnarray}
where
\begin{equation}
\boldsymbol{S}' = \frac{1}{2} \left( \mbox{grad} \, \boldsymbol{u}' + (\mbox{grad} \, \boldsymbol{u}')^{\mathrm{T}}  \right)
\end{equation}
is the strain rate tensor. 
Assuming the medium as a calorically perfect gas, the set of equations is closed by the ideal gas equation of state
\begin{equation}
p = \frac{\gamma - 1}{\gamma} \rho T \mbox{.}
\end{equation}

\bibliographystyle{plainnat}
\bibliography{biblio}

\end{document}